\documentclass[3p]{elsarticle}
\usepackage{graphicx}
\usepackage{epsfig}
\usepackage{amssymb}
\usepackage{amsthm}
\usepackage{multirow}
\usepackage[table]{xcolor}
\usepackage{float}
\usepackage{dblfloatfix}
\usepackage{soul}
\usepackage[normalem]{ulem}
\usepackage{url}
\usepackage{hyperref}
\usepackage{booktabs}
\usepackage{caption} 
\usepackage{subcaption}      
\usepackage{siunitx}    
\usepackage{comment}
\usepackage{longtable}
\usepackage{dirtree}

\newcommand{\cmark}{\ding{51}}%
\newcommand{\xmark}{\ding{55}}%
\definecolor{Gray}{gray}{0.9}
\definecolor{maroon}{cmyk}{0,0.87,0.68,0.32}
\begin{document}

\begin{frontmatter}

\title{LibriVAD: A Scalable Open Dataset with  Deep Learning Benchmarks for Robust Voice Activity Detection}

\author{Ioannis Stylianou$^{a, b,}$\fnref{cor1}}
\address[]{Department of Electronic Systems,
Aalborg University, Denmark \\
 {\small \tt }}
 \address[]{Pioneer Centre for AI, Denmark  \\
 {\small \tt }}
\author{Achintya Kr. Sarkar$^{c,}$\fnref{cor1}}
 \address[]{ECE Group, IIIT SriCity, Chittoor, India\\
 {\small \tt }}
 \author{Nauman Dawalatabad$^{d,}$\fnref{cor2}}
 \address[]{ Zoom Communications Inc., USA\\
 {\small \tt }}
 \author{James Glass$^e$}
 \address[]{Computer Science and Artificial Intelligence Laboratory, Massachusetts Institute of Technology, Cambridge MA, USA \\
 {\small \tt }}
 \author{Zheng-Hua Tan$^{a, b}$\corref{cor3}}

\fntext[cor1]{Equal contribution}

\fntext[cor2]{This work was done while at MIT Computer Science and Artificial Intelligence Laboratory, USA.}

\cortext[cor3]{Corresponding author. Email address: zt@es.aau.dk}

\date{November 2025}

\begin{abstract}
Robust Voice Activity Detection (VAD) remains a challenging task, especially under noisy, diverse, and unseen acoustic conditions. Beyond algorithmic development, a key limitation in advancing VAD research is the lack of large-scale, systematically controlled, and publicly available datasets. 
To address this, we introduce \textbf{LibriVAD} -- a scalable open-source dataset derived from LibriSpeech and augmented with diverse real-world and synthetic noise sources. LibriVAD enables systematic control over speech-to-noise ratio, silence-to-speech ratio (SSR), and noise diversity, and is released in three sizes (15 GB, 150 GB, and 1.5 TB) with two variants (LibriVAD-NonConcat and LibriVAD-Concat) to support different experimental setups.
We benchmark multiple feature-model combinations, including  waveform, Mel-Frequency
Cepstral Coefficients (MFCC), and Gammatone filter bank cepstral coefficients, and introduce the  Vision Transformer (ViT) architecture for VAD. Our experiments show that ViT with MFCC features consistently outperforms established VAD models such as boosted deep neural network and convolutional long short-term memory deep neural network across seen, unseen,
and out-of-distribution (OOD) conditions, including evaluation on the real-world VOiCES dataset. 
We further analyze the impact of dataset size and SSR on model generalization. Experimental results show that scaling the dataset size from small-scale to medium- or large-scale noticeably and consistently improves VAD performance under OOD conditions, while appropriately balancing SSR enhances OOD performance in the small-scale setting.
All datasets, trained models, and code are publicly available to foster reproducibility and accelerate progress in VAD research.
\end{abstract}

\end{frontmatter}

\section{Introduction} \label{introduction}

Voice Activity Detection (VAD) is a fundamental technology in speech processing that aims to distinguish between speech and non-speech segments in audio signals \cite{tan2020rvad}. It plays a critical role in a wide range of applications, including automatic speech recognition \cite{10023187,9747357}, speaker identification \cite{HOANG2025104969}, keyword spotting \cite{sarkar2023,10890730,10858118}, speaker diarization \cite{Fujita2019, Horiguchi2020, 10890295}, speech enhancement \cite{hoang2022minimum}, and blind speech separation \cite{Opochinsky2025, 10888445}. By filtering out non-speech audio segments, VAD improves both the efficiency and robustness of these systems.
VAD methods typically consist of two main components: a signal representation--such as Mel-Frequency Cepstral Coefficients (MFCC) \cite{Davis80}, Gammatone Filter bank Cepstral Coefficients (GFCC) \cite{Valero,4517928}, or raw waveform \cite{zazo2016feature}--and a classification model, which can be supervised \cite{zazo2016feature,7347379,Wave_2020} or unsupervised \cite{tan2020rvad, Sohn1999a}.

Although numerous methods have been proposed, their evaluation 
has remained limited by the lack of large‑scale, publicly available, and acoustically diverse datasets. Existing corpora often exhibit constraints such as restricted scale, domain specificity, limited noise variability, or licensing restrictions. 
For example, the RATS (Radio Traffic Collection System) dataset \cite{Walker2012} was designed for radio communication and includes channel distortions and burst-like noise, which are not representative of typical everyday environments. 
TIMIT \cite{Timit} (and its noisy variant \cite{dean10_interspeech}, N-TIMIT \cite{NTimit})
and AURORA-2 \cite{Hirsch2000}  
are not publicly available and offer limited noise diversity. Additionally, their utterances share a fixed temporal structure -- short silences at the beginning and end, with speech in the middle -- causing VAD systems to exploit this pattern and potentially overfit to these structural biases \cite{larsen2022adversarial}. The AVA-Speech dataset \cite{chaudhuri18_interspeech}, constructed from YouTube movie videos, offers more realistic content but still provides limited noise coverage. WSJ0  \cite{LDC93S6A} has also been employed for VAD in a noise-robust study \cite{li2024robust}, where labels were generated by applying the open-source rVAD \cite{tan2020rvad} to clean speech prior to noise mixing; however, WSJ0 itself is not publicly available.
The Fearless Steps Challenge \cite{fearlessstepschallengephase1} focuses on historical mission audio (Apollo) with unique acoustic properties, limiting its relevance to modern applications. Collectively, the absence of a large, publicly available, and systematically controlled VAD benchmark hinders fair comparison and advancement of VAD algorithms. 

In contrast, other areas of speech and audio processing benefit from standardized benchmarks, such as NIST SRE for speaker recognition \cite{SRE}, LRE for language recognition \cite{LRE}, ESC \cite{piczak2015dataset} for environmental sound classification, DCASE \cite{heittola2020acousticsceneclassificationdcase} for acoustic scene and sound classification, and LibriMix for speech separation \cite{cosentino2020librimix}.

To address this gap, we introduce \textbf{LibriVAD}: a large-scale, open-source dataset derived from the LibriSpeech corpus \cite{kahn2020libri} and enriched with a wide range of real-world and synthetic noise sources. LibriVAD is designed to support evaluation of VAD systems under challenging and systematically varied acoustic conditions through controled Signal-to-Noise  Ratio (SNR) and Silence-to-Speech Ratio (SSR) settings, and through diverse real-world noise types drawn WHAM! \cite{wichern2019wham} and DEMAND \cite{thiemann_2018_1227121}, and artificially generated babble and speech-shaped noise. Beyond VAD, LibriVAD can also serve as a resource for speech processing tasks such as speech enhancement, noise-robust training, and general speech-noise mixture modeling.

The dataset is released in three scalable sizes (15GB, 150GB, and 1.5TB) to accommodate different computational budgets.
It further includes two variants: LibriVAD-NonConcat, which resembles commonly used datasets with limited non-speech content, and LibriVAD-Concat, which contains a higher proportion of non-speech segments to better reflect real-world environments where speech occurs intermittently. 
Furthermore, LibriVAD enables evaluation under both seen and unseen conditions through its dedicated test sets. To assess out-of-distribution performance, we use the VOiCES dataset \cite{richey2018voices}, which includes audio recordings with strong reverberation--absent in the LibriVAD training data--and real-world additive noise. While LibriSpeech has previously been used for VAD research (e.g., personalized VAD in \cite{ding2019personal}, \cite{bovbjerg2024self}), these earlier works rely on concatenating single-speaker utterances from different speakers to simulate multi-speaker scenarios, limiting their suitability as general-purpose VAD benchmarks.



In addition to introducing LibriVAD, we apply the Vision Transformer (ViT) \cite{dosovitskiy2020image} architecture to VAD for the first time, leveraging its strong ability to model long-range dependencies and contextual cues. We evaluate ViT using MFCC and GFCC features and compare its performance against established VAD models including Silero \cite{SileroVAD}, Boosted Deep Neural Network (bDNN) \cite{7347379} and Convolutional Long Short-Term Deep Neural Network (CLDNN) \cite{zazo2016feature}.

Our key contributions are as follows:

\begin{itemize}
\item \textbf{Dataset Creation}: We introduce LibriVAD, a large-scale dataset comprising speech-noise mixtures across multiple SNR levels, SSR values, and diverse noise types. 

\item \textbf{Baseline Evaluation}: We benchmark several VAD feature-model combinations, providing strong and reproducible baselines.

\item \textbf{Open-Source Release}: To foster reproducibility and accelerate progress in VAD research and applications, we publicly release the entire dataset\footnote{\url{https://huggingface.co/datasets/LibriVAD/LibriVAD}}, baseline implementations, code, and trained models\footnote{\url{https://github.com/zhenghuatan/librivad}}. 

\item \textbf{Empirical Insights}: We provide an in-depth analysis of how dataset size, SSR and model size influence VAD model performance and generalization.

\item \textbf{Novel VAD Model}: We develop and evaluate a ViT-based VAD approach and demonstrate its superior performance compared to existing models.
\end{itemize}


The remainder of this paper is organized as follows: Section \ref{LibriVAD} presents the core components of LibriVAD and the construction process. Section \ref{Methods} outlines the feature extraction methods, model architectures, evaluation metrics, and experimental setup. Section \ref{Results} presents the performance of all the proposed and baseline methods under different conditions and datasets, provides a statistical analysis of the results, and discusses the limitations of the paper. Finally, Section \ref{Conclusion} concludes the paper.

\section{LibriVAD} \label{LibriVAD}
LibriVAD is a large-scale, open-source dataset that combines clean LibriSpeech audio with diverse noise sources and silent patterns to support VAD training and evaluation. 

\subsection{Silence Injection and Silence-to-Speech Ratio Adjustment} \label{subsec:ssr}
To study the impact of silence distribution on VAD performance, we control the portion of silence in the dataset by constructing two variants of clean speech:
\begin{itemize}
    
   \item {\bf LibriSpeech \cite{panayotov2015librispeech}}. As one of the most popular open-source corpus of English speech, LibriSpeech is derived from LibriVox audiobooks, and is widely used for automatic speech recognition and other speech-related research. It comprises approximately 1,000 hours of speech data with high-quality transcriptions and features a diverse set of speakers. 
   In this work, we use the following clean subsets of the LibriSpeech dataset; \texttt{train-clean-100}, \texttt{test-clean} and \texttt{dev-clean}. 

    \item {\bf LibriSpeech-Concat:} In the original LibriSpeech dataset, silence accounts for approximately 17.6\% of the total audio. Since VAD is a binary classification task (speech vs. non-speech), this imbalance can affect supervised learning performance. 
    To address this, we construct \texttt{LibriSpeech-Concat}, where each entry is formed by concatenating two consecutive LibriSpeech utterances and inserting silence between them. The inserted silence is calibrated to be 25\% of the combined duration of the two signals, resulting in a Silence-to-Speech Ratio (SSR) of approximately 34\%, where
    \begin{equation}
        \text{SSR} = \frac{\text{Duration of Silence}}{\text{Duration of Speech}}
    \end{equation}
    This chosen ratio is not arbitrary; rather, it is deliberately selected to align the dataset's temporal characteristics with those of natural, spontaneous communication. While read speech (such as the LibriSpeech corpus) typically exhibits low silence ratios due to the absence of cognitive planning pauses \cite{henderson1966sequential}, spontaneous human speech contains significantly higher proportions of silence. It is estimated \cite{goldman1968psycholinguistics} that silence can account for between 30\% and 50\% of the total duration in spontaneous utterances, driven by hesitation and lexical formulation. By elevating the SSR to approximately 34\%, we shift the temporal profile of \texttt{LibriSpeech-Concat} from the domain of read speech toward the statistical profile of conversational speech, thereby providing a more realistic challenge for robust VAD systems.

    To achieve this, we utilize publicly available forced alignments \cite{Jemine2017LibrispeechAlignments}, extract the silent segments from the \texttt{train-clean-100} subset of LibriSpeech, and concatenate them to form a unified silence signal of approximately 17.5 hours. This silence signal is then segmented sequentially; starting from its beginning, a silence chunk of the required duration is extracted and inserted between each pair of consecutive LibriSpeech files, after which the process continues from the remaining portion of the signal without reuse. When the end of the silence signal is reached, the remaining samples are used and the segmentation process restarts from the beginning of the signal, effectively wrapping around and continuing sequentially.
\end{itemize} 

Table \ref{table:LibriconcatNo} provides a summary of the two datasets, including signal lengths, speech durations, and the number of signals.

\begin{figure}[H]
    \centering
    \includegraphics[width=\linewidth]{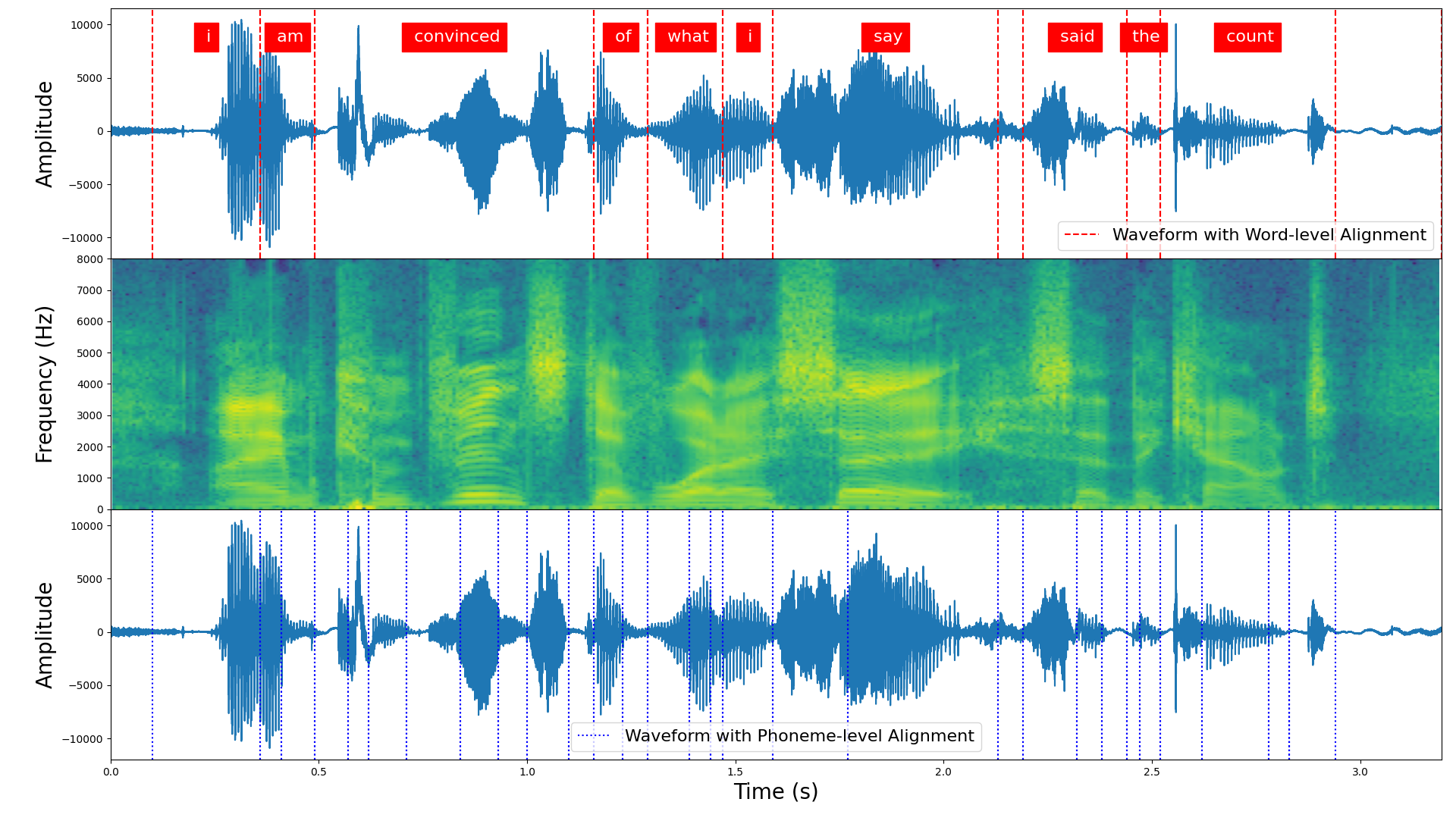}
    \caption{An example of forced alignments on a clip from the LibriSpeech dataset. The bottom plot illustrates the phoneme boundaries, the top plot shows the corresponding word boundaries generated by the Montreal Forced Aligner, and the middle plot presents the spectrogram.}
    \label{fig:forced_alignments}
\end{figure}

Since  forced alignments can reliably distinguish between speech and non-speech, we also use them to generate frame-level labels for the VAD task, as exemplified in Figure \ref{fig:forced_alignments}. Forced alignments obtained via automatic speech recognition have been shown to produce accurate and consistent VAD labels--even comparable to those provided by an expert labeler \cite{kraljevski2015comparison}. 

\begin{table}[h]
\caption{\it Characteristics of LibriSpeech and LibriSpeech-Concat datasets. 
}
\resizebox{\textwidth}{!}{
\begin{tabular}{|l|l|c|c|c|c|c|c|}  \hline 
     \multicolumn{2}{|c|}{} & \multicolumn{6}{|c|}{ Datasets}        \\ 
     \multicolumn{2}{|c|}{} &  \multicolumn{3}{|c|}{LibriSpeech}   &  \multicolumn{3}{|c|}{ LibriSpeech-Concat } \\ 
 Category & Characteristics & train-clean-100& test-clean& dev-clean & train-clean-100& test-clean& dev-clean\\ \hline
 \hline
 \multirow{5}{*}{Signal} & Average length & 12.71s& 7.42s& 7.18s & 31.72s& 18.56s& 17.94s\\
 & Standard Deviation & 3.56s& 5.15s&4.69s & 6.80s& 10.31s& 9.29s\\ 
 & Maximum length & 24.53s& 34.96s& 32.65s &  45.70s& 69.29s& 69.76s\\
 & Minimum length & 1.41s& 1.29s& 1.45s & 5.32s& 4.84s& 5.05s\\
 & Total duration & 100.74h & 5.40h & 5.39h & 125.72h & 6.75h & 6.73h \\
  \hline
 \multirow{5}{*}{Speech} & Average length & 10.51s& 6.21s& 5.97s & 20.95s& 12.41s& 11.94s\\
 & Standard Deviation &3.07s& 4.47s& 4.14s & 4.76s& 7.13s& 6.58s\\
 & Maximum length & 23.50s&30.29s& 30.21s & 34.27s& 45s& 49.83s\\
 & Minimum length & 0.76s& 0.69s& 0.7s & 3.02s& 2.37s& 2.54s\\
 & Total duration & 83.31h & 4.52h & 4.48h & 83.03h & 4.52h & 4.48h \\
 \hline
 \hline
 \multicolumn{2}{|l|}{Number of signals} &28535& 2620&2703 &14267&1310& 1351\\
 \hline
\end{tabular}
}
\label{table:LibriconcatNo}
\end{table}

\subsection{Noise Selection and Generation} \label{Noise}

To simulate realistic acoustic environments, LibriVAD incorporates a diverse range of noise sources, including both publicly available datasets and our custom-generated noises that are also released for public use. These noise types span a wide range of everyday scenarios.

\begin{description}
\item{\textbf{Public noise datasets:}}
\begin{itemize}

    \item \textbf{WHAM!} \cite{wichern2019wham}: This dataset was recorded in urban environments (including restaurants, cafes, bars, and parks) 
using binaural microphones, and is utilized to create the ``city" noise category. 
Table \ref{table:WHAM dataset} summarizes the details of the WHAM! dataset.
The WHAM! noise dataset provides default splits for training, validation, and testing. 
Due to its large size, we use only a representative subset. For each split, signals are selected and concatenated to preserve the distribution of recording environments. The resulting durations for the training, validation, and test splits are 3 hours, 30 minutes, and 30 minutes, respectively. 

\begin{table}[h]
\caption{\footnotesize Signal length statistics of the WHAM! dataset.}
\center
\scalebox{1.1}{
\begin{tabular}{ |p{4.2cm}|p{1.6cm}| }
 \hline
 Average signal length &10s\\
 \hline
 Maximum signal length &47.7s\\
 \hline
 Minimum signal length &3.5s\\
 \hline
 Number of signals &28000\\
 \hline
\end{tabular}}
\label{table:WHAM dataset}
\end{table}

    \item \textbf{DEMAND } \cite{thiemann_2018_1227121}: This database categorizes six environments into indoor and outdoor settings, each represented by three recordings, as shown in Table \ref{table:Demanddatasets}.  
    Each recording consists of sixteen 15-minute audio segments (1–16), grouped as 1–12 for training, 13–14 for validation, and 15–16 for testing. Segments from all environments are then concatenated to form the final splits for six noise types.
\end{itemize}

\begin{table*}[!t]
\caption{Overview of the different recording environments and conditions available in the DEMAND noise dataset.} 
\resizebox{\textwidth}{!}{
\begin{tabular}{|c|l|}\cline{1-2} 
Environments  &  \multicolumn{1}{c|}{Conditions}  \\ \hline
        &\textbf{Domestic:}         Washroom (DWASHING), Kitchen (DKITCHEN), Living Room (DLIVING) \\
Indoor  & \textbf{Office:}          Office (OOFFICE), Hallway (OHALLWAY),  Meeting Room (OMEETING) \\
        &\textbf{Public:}           Subway Station (PSTATION), Cafeteria (PCAFETER),  Restaurant (PRESTO) \\
        & \textbf{Transportation:}   Subway (TMETRO), Bus (TBUS),Car (TCAR) \\ \hline
Outdoor  &    \textbf{Nature:}      Field (NFIELD), River (NRIVER), Park (NPARK) \\
         & \textbf{Street:}         Traffic Intersection (STRAFFIC), Town Square (SPSQUARE),  Cafe Terrace (SCAFE) \\ \hline
        \end{tabular}
}
\label{table:Demanddatasets}
\end{table*}

\item{\textbf{Synthetic noise sources, which is made publicly available\footnote{\url{https://huggingface.co/datasets/LibriVAD/LibriVAD}}:}} 


\begin{itemize}

\item{\textbf{Speech Shaped Noise (SSN)}}: This noise source is generated using the LibriLight-small dataset \cite{kahn2020libri}, which was selected to ensure it is completely separate from LibriSpeech. First, audio signals are grouped by speaker identity. For each speaker, recordings are selected based on two criteria: 
an SNR above 15 dB and an SSR exceeding 90\%. Speakers with more than 10 minutes of speech after selection are included for SSN generation.
In total, 22 speakers satisfy these criteria. Each selected speaker’s audio is analyzed using 12th-order linear predictive coding, producing coefficients that are used to construct an all-pole filter. Applying this filter to Gaussian white noise yields a signal that matches the speaker’s long-term average speech spectrum, generating “personalized” SSN for each individual, providing more realistic and challenging noise for evaluating VAD systems.

To generate the training, validation, and testing SSN signals, filters derived from the first 18 speakers are used for training, the subsequent two for validation, and the final two for testing. For training, we extract 200 seconds of signal from each of the first 18 filters and concatenate these to form a one-hour signal. Similarly, we extract 300 seconds of signal from each validation and test filter, resulting in 10-minute signals for each split. 

    \item{\textbf{Babble Noise}}: This noise source is generated using the LibriLight-medium dataset \cite{kahn2020libri}, which was chosen to maintain dataset separation from LibriSpeech. Similar to SSN generation process, only utterances with an SNR above 15 dB are retained. In addition, we apply the open-source rVAD technique \cite{tan2020rvad} to detect and remove silent sections. The Babble noise is created by mixing six energy-normalized speech streams into a single combined signal. Each stream is formed by concatenating the selected utterances in sequence, ensuring that each speaker is assigned to only one stream. Removing silence prevents simultaneous pauses across multiple streams, resulting in a smoother and more consistent noise signal. 
\end{itemize}

\end{description}

Table \ref{table:sumnoise} summarizes the characteristics of the generated noises.

\begin{table}[h]
\caption{Overview of the noise sources used in LibriVAD, including noise type, source, duration of each split, and whether the noise is used for generating the training set (seen) or not (unseen). Training splits for unseen noises are still included, so researchers can experiment with different seen-unseen combinations.}
\centering 
\scalebox{1}{
\begin{tabular}{ |p{2.8cm}|p{3.2cm}|p{3cm}|p{1.2cm}|p{.7cm}|}
 \hline
 \multicolumn{5}{|c|}{Noise Specifics} \\
 \hline
 Noise type &Source&Duration (Train,Val,Test)&Unseen&Seen\\
 \hline
 \hline
 Babble & LibriLight-medium& (3h, 30m, 30m)& \cmark& \xmark\\
 \hline
 Speech Shaped &LibriLight-small& (1h, 10m, 10m)& \cmark& \xmark\\
 \hline
 Domestic  &DEMAND& (3h, 30m, 30m)& \cmark& \xmark\\
 \hline
 \hline
 Nature  &DEMAND& (3h, 30m, 30m)& \xmark& \cmark\\
 \hline
 Office  &DEMAND& (3h, 30m, 30m)& \xmark& \cmark\\
 \hline
 Public  &DEMAND& (3h, 30m, 30m)& \xmark& \cmark\\
 \hline
 Street  &DEMAND& (3h, 30m, 30m)& \xmark& \cmark\\
 \hline
 Transport  &DEMAND& (3h, 30m, 30m)& \xmark& \cmark\\
 \hline
 City  &WHAM!&(3h, 30m, 30m)& \xmark& \cmark\\
 \hline
\end{tabular}}
\label{table:sumnoise}
\end{table}

\subsection{LibriVAD Dataset Composition and Generation Process}\label{subsec:final-dataset}

Entries from both the LibriSpeech and LibriSpeech-Concat datasets are mixed with nine noise types (City, Domestic, Office, Public, Transport, Nature, Street, Speech Shaped Noise, and Babble) at six SNR levels (-5, 0, 5, 10, 15 and 20 dB). SNR is computed using only speech-active segments to ensure a more accurate SNR measure, as including silent regions can underestimate the scaling of the noise level.  More specifically, for a given clean speech signal $x(t)$, an equal-in-length noise sub-segment $n(t)$ is iteratively selected and scaled by $\lambda$ to satisfy the target Signal-to-Noise Ratio (SNR) according to

\begin{equation}
\text{SNR}_{\mathrm{dB}} = 20 \log_{10} \left( \frac{A_{\mathrm{speech}}}{\lambda A_{\mathrm{noise}}} \right) \end{equation}
which yields  
\begin{equation}
\lambda = \frac{A_{\mathrm{speech}}}{A_{\mathrm{noise}}} \cdot 10^{-\frac{\text{SNR}_{\mathrm{dB}}}{20}}
\end{equation}

\noindent where $A$ denotes the Root Mean Square (RMS) amplitude.\\

The two signals are then superimposed to generate the corrupted file at the specified SNR level. If the full noise file is exhausted and the sub-segment is shorter than the speech signal, the sub-segment is cyclically extended (restarting from the beginning of the full noise signal) to cover the duration of the utterance.

After corruption at all SNRs and Noises, the resulting datasets are referred to as \emph{LibriVAD-NonConcat} and \emph{LibriVAD-Concat}, corresponding to their respective clean speech counterparts. The two variants differ in temporal structure. \emph{LibriVAD-Concat} forms longer continuous sequences, enabling models to learn temporal dependencies across utterance boundaries. \emph{LibriVAD-NonConcat}, in contrast, retains strict utterance-level segmentation.

To enhance accessibility and accommodate varying computational resources, the LibriVAD dataset is offered in three variants: large, medium, and small. The large variant comprises the full, unabridged 1.5 TB corpus, encompassing roughly 14,000 hours of audio with all generated noisy utterances. The medium (150 GB / 1,400 h) and small (15 GB / 140 h) variants are created via systematic sub-sampling of the alphanumerically sorted clean source files--retaining every 10th file for the medium set and every 100th file for the small set. These logarithmic intervals were selected to reflect practical research constraints: The Small subset fits within the memory limits of consumer-grade hardware, enabling rapid architectural prototyping; the Medium subset strikes a balance for standard academic workstations with multi-GPU setups; and the Large scale addresses the need for greater data volume and stronger benchmarking capacity in high-compute environments. The deterministic sub-sampling approach preserves the distributional characteristics of the full dataset, ensuring that speaker diversity, acoustic conditions, noise types, and SNR levels are maintained, making the smaller subsets representative proxies for large-scale experiments. Table \ref{table:libriVAD_summary} summarizes the different size characteristics of the LibriVAD dataset. 

 To position LibriVAD relative to existing VAD datasets, Table \ref{table:otherdataset}  provides a comparative overview, highlighting differences in scalability, noise characteristics, and availability. Note that proprietary datasets are also commonly used in the literature (e.g., \cite{Wave_2020}), but these are not included in our comparison.

\begin{table}[h]
\centering
\caption{Overview of LibriVAD dataset variants, including their respective sizes and durations.}
\label{table:libriVAD_summary}
\renewcommand{\arraystretch}{1.2}
\scalebox{0.95}{
\begin{tabular}{|l|c|c|c|}
\hline
\multicolumn{4}{|c|}{\textbf{LibriVAD Dataset Variants}} \\ \hline
\textbf{Dataset Variant} & \textbf{Size Variant} & \textbf{Size (Train, Val, Test)} & \textbf{Duration (Train, Val, Test)} \\ \hline

\multirow{3}{*}{\textbf{LibriVAD-NonConcat}} 
& Small  & 5.9GB, 0.34GB, 0.33GB& 54.6, 3.1, 3.1 hours\\ 
& Medium & 59GB, 3GB, 3.3GB & 545.9, 27.4, 30.2 hours \\
& Large  & 584GB, 32GB, 32GB & 5440, 291, 292 hours  \\ \hline

\multirow{3}{*}{\textbf{LibriVAD-Concat}} 
& Small  & 7.4GB, 0.39GB, 0.45GB & 68.7, 3.5, 4.1 hours\\ 
& Medium & 73GB, 3.8GB, 3.9GB & 679.8, 35.2, 35.9 hours \\ 
& Large  & 729GB, 40GB, 40GB & 6788, 364, 365 hours\\ \hline

\end{tabular}}
\end{table}

\begin{table}[t]
\centering
\caption{Comparison of LibriVAD and existing VAD datasets across duration, noise characteristics, and availability.}
\begin{tabular}{|p{2cm}|p{2cm}|p{1.5cm}|p{3.5cm}| p{1.7cm}| p{2cm}|}
\hline
\textbf{Dataset}    &   \textbf{Source Duration (hrs)} & \textbf{Full Duration (hrs)}& \textbf{Noise types and data characteristics} &  \textbf{SNR (dB)} &  \textbf{Availability} \\ \hline \hline
AVA-Speech \cite{chaudhuri18_interspeech}     &Unspecified & $\sim 45$& Speech with background music/noise from movie audio; speaker count unspecified &        Unspecified and unchangeable&  Open source\\ \hline
RATs \cite{Walker2012}     & 350 &  3000&  Channel distortion; 16kHz; speaker count unspecified           &            Unspecified&  Purchased license \\ \hline
Apollo \cite{hansen19_interspeech}       & - &  80       & Channel distortion  and extended periods of non-speech; 8kHz; speaker count unspecified &    0-20 (approx.) & Open source\\ \hline
Aurora-2 \cite{Hirsch2000}    & $\sim 6$  & $\sim 39$&Airport, babble, car, exhibition, restaurant, street, subway, train; 8kHz; 110 speakers&           -5, 0, 5,  10, 15, 20      & Purchased license \\   \hline           
N-Timit \cite{NTimit}   & 5  &  $\sim 55$    &        Telephone‑channel distortion with narrowband characteristics; stored at 16kHz; 630 speakers.  & Unspecified    &     Purchased license \\ \hline
LibriVAD-NonConcat   &   100 &       $55$ / $546$ / $5440$   &      Nature, office, public, street, transport, city, babble, SSN, domestic; 16kHz; 330 speakers &   -5, 0, 5,  10, 15, 20  & Open source (MIT)\\ \hline 
LibriVAD-Concat      &   100 &       $ 69$ / $680,$ / $6788$ &  Same as NonConcat, with concatenated sequences  & -5, 0, 5,  10, 15, 20     & Open source (MIT)\\
\hline
\end{tabular}
\label{table:otherdataset}
\end{table}

\section{VAD Methods} \label{Methods}
This section describes the methods used to benchmark the generated datasets, starting with the various spectral presentations of signals used in VAD classification.

\subsection{Features}

\begin{itemize}
    
\item{\textbf{Gammatone filter bank cepstral coefficients \cite{Valero}:}}
The GFCC features of $39$ dimensions ($12$ cepstral coefficients plus log energy, along with their first- and second-order derivatives, $\Delta$ and $\Delta \Delta$). Features are extracted using a $25ms$ window and a $10ms$ frame shift. The gammatone filter bank spans the frequency range $[50, \text{sampling-frequency}/2]$ with the number of filters determined by $\text{hz2erb}(\text{sampling frequency}/2) - \text{hz2erb}(50)$
, where hz2erb denotes the equivalent rectangular bandwidth (ERB) scale. This results in approximately one filter every $0.9mm$ along the cochlea. Finally, feature vectors are normalized to zero mean and unit variance at the utterance level. The GFCC implementation is from \cite{yu2017spoofing}.

\item{\textbf{Mel-frequency cepstral coefficients \cite{Davis80}:}}
MFCC features also comprise $39$ dimensions (coefficients $C_{1-12}$ plus log energy, and their $\Delta$ and $\Delta \Delta$ derivatives). They are extracted using $24$ Mel filters, a $25ms$ Hamming window, and a $10ms$ frame shift. As with GFCC, feature vectors are normalized to zero mean and unit variance per utterance. 

\end{itemize}

\subsection{VAD Models}
In this subsection, we present the VAD methods used to establish baseline performance, employing various deep learning models.

\textbf{raw-CLDNN}:
The convolutional long short-term memory deep neural network (CLDNN) \cite{zazo2016feature} processes directly raw audio waveforms. It combines convolutional layers for feature learning, LSTM layers for modeling temporal dependencies, and fully connected layers for classification. The architecture includes two convolutional layers, three unidirectional LSTM layers, and one feed-forward layer, closely approximating the original model with approximately 200k parameters.

\textbf{bDNN}:
In this method \cite{7347379}, the input feature vector and its corresponding label for the current frame are expanded by incorporating neighboring frames to provide contextual information.
The resulting expanded feature vector is fed into a DNN, which maps the input to a set of output nodes. The number of output nodes equals to the total number of frames considered after adding the contextual neighbors.  The bDNN architecture consists of two hidden feed-forward layers (excluding the input and output layers), each containing $512$ neurons. A sigmoid activation function is applied to the output layer to produce values between 0 and 1.
The network parameters are optimized using the  $L_1$ loss function, minimizing the difference between the original and predicted labels for both the current and contextual frames. During inference, the outputs from all nodes are averaged to compute the probability that the current frame belongs to the speech class.

\textbf{Silero}: This is an enterprise-grade VAD method developed by the Silero team  \cite{SileroVAD}.   It is a lightweight neural network model designed for both online and offline  VAD.  While the model binaries are publicly available, the internal model architecture, such as layer definitions, training procedures, and other implementation details, has not been disclosed.

\textbf{ViT}:
We employ the keyword transformer (KWT-3) \cite{Berg_2021}, which adapts  the ViT \cite{dosovitskiy2020image}, for the VAD task. 
Feature vectors are split into sequences of $100$ frames and passed through patching, positional encoding, and several transformer encoder blocks. The final encoder output is projected onto two output nodes corresponding to speech and non-speech classes at the frame level. For sequences or residual segments shorter than 100 frames, zero-padding is applied to ensure a consistent input length. 

It is well established that ViT can model long-range dependencies \cite{khan2022transformers}. Beyond this,
ViT offers several properties particularly advantageous for VAD. Its self-attention mechanism captures both local and global acoustic cues and adaptively down-weights noise-
dominated patches, thus improving robustness
under nonstationary conditions such as babble noise. In addition, the frame-level patch formulation
in KWT-3 \cite{Berg_2021} facilitates direct frame-level VAD decisions,
making it well suitable for frame-level VAD tasks. This design allows ViT to focus on informative temporal regions, such as short speech bursts and pauses, whereas
CLDNN models rely more heavily on local transitions due to their fixed local receptive fields, and bDNN operates on a fixed and limited input window without explicit temporal modeling. Conseqently, both architectures lack the ability of ViT to capture long-term dependencies in the audio signal.

\subsection{Evaluation Metrics} \label{Metric}


\textbf{Area under the curve (AUC)}:
The AUC metric measures the area under the Receiver Operating Characteristic (ROC) curve, which plots the true positive rate (TPR) and false positive rate (FPR) at various thresholds: 
\begin{eqnarray}
\text{TPR} & = & \frac{TP}{TP + FN} \\
\text{FPR} & = & \frac{FP}{FP + TN}
\end{eqnarray}
where  $TP$ = True Positives, $TN$ = True Negatives,  $FP$ = False Positives, and  $FN$ = False Negatives.\\

\textbf{Equal error rate (EER) and Minimum detection cost function (MinDCF)}: EER represents the operating point where the miss rate (a speech frame incorrectly identified as non-speech) equals to the false acceptance rate (a non-speech frame incorrectly identified as speech). MinDCF incorporates application-specific penalties for missed targets and false acceptances, providing a weighted trade-off between these  two types of errors. The Detection cost  function (DCF) is defined as:  

 \begin{eqnarray}
    DCF & = &  C_{miss} \times P_{miss|target} \times P_{target} \\ \nonumber
    & & + C_{FA} \times P_{FA | non-target} \times (1-P_{target}) 
 \end{eqnarray}
 where the $C_{miss}=10, C_{FA}=1, \text{and } P_{target}=0.01$ indicate the cost of a miss, the cost of a false alarm, and the a priori probability of the target, respectively, as per NIST 2008 SRE evaluation plan. Both EER and MinDCF are commonly used in NIST speaker recognition evaluation \cite{SRE, OpenSAD15}.

\subsection{Experimental Setup}
For MFCC and GFCC feature extraction, we use $24$ Mel-frequency filter banks and $64$ Gammatone filter banks, respectively. A pre-emphasis coefficient of $0.97$ is applied.  
For the raw-CLDNN model, raw audio signals are segmented using a $35ms$ window with a $10ms$ frame shift, following the setup in \cite{zazo2016feature}. The segmented audio signal is then passed through time and frequency convolution layers, followed by three LSTM layers. The final output is projected onto classification nodes representing speech and non-speech classes. 
In the bDNN model, the contextual parameters are set to $u=9$ (left-context frames) and $w=19$ (right-context frames). Both the CLDNN and bDNN models are trained using Stochastic Gradient Descent (SGD), with a learning rate $0.001$ and a batch size of $512$. The result  reported in Tables \ref{tab:LibriSpeech-NonConcat-small} and \ref{tab:LibriSpeech-Concat-small} are based on this training configuration. To verify that the performance differences were not caused by optimization setting, we additionally re-implemented and retrained the bDNN and CLDNN models using the same AdamW optimizer, learning rate, and warm-up strategy  as ViT‑MFCC model. These sanity-check experiments yielded results closely matching those reported in Tables \ref{tab:LibriSpeech-NonConcat-small} and \ref{tab:LibriSpeech-Concat-small}.

In the ViT model, each input frame is treated as a patch of size \texttt{[1, \text{feature dimension}]} and is projected onto a $192$-dimensional embedding space.  The ViT architecture consists of $12$ encoder layers, each employing a $3$-head multi-head attention mechanism with $64$ dimensions per head. The model is optimized using the AdamW optimizer (with a weight decay of $0.1$). The MLP-layer includes layer normalization followed by a feedforward network that expands the $192$-dimensional input to $768$ dimensions, which is then projected onto two output nodes representing speech and non-speech classes. Batch size is set to $32$. All DNN models are trained for up to $50$ epochs  from scratch, and no early-stopping criterion is applied. For the ViT, each input sequence is divided into $100$ time steps (zero-padded for shorter sequences), and the learning rate ($0.001$) is scheduled using cosine-warmup annealing over the first $10$ epochs. The implementation of the  ViT is adopted from \cite{smith2024mycode}. 

 For out-of-distribution evaluation, we use the VOiCES DevKit dataset as in \cite{richey2018voices}. This dataset is a subset of the VOiCES dataset, where distracted audio files are generated by simultaneously playing clean speech and  different distractor noises (television, radio, babble, music) and recording them using microphones (cardioid dynamic studio microphone and omnidirectional condenser lavalier microphones, including various microphone sets) placed at both close and far distances in rooms (acoustic conditions including reverberation and HVAC noise) of four different sizes (146'' × 107'', 225'' × 158'',  25' x 25', 32' x 16' plus 10' x 10' ). This setup captures both reverberation and spatial variability. The dataset contains a total of $19,200$ audio files. 
 
 To obtain ground-truth VAD labels for the VOiCES dataset, we employ the open-source Montreal Forced Aligner \cite{mcauliffe17_interspeech} (with pre-trained English\_us\_arpa acoustic and language models) on the original clean LibriSpeech recordings (prior to playback) using the corresponding transcripts, as commonly done in the literature. These VAD labels are then reused for the corresponding noisy or distracted versions of the same recordings. We provide \textit{the generated VAD labels for the VOiCES dataset}  as part of the LibriVAD release. 
 
 The speakers in the VOiCES devkit dataset are disjoint from those in our training datasets (LibriSpeech-NonConcat and LibriSpeech-Concat), ensuring that the evaluation truly measures how well the model's ability to generalize to unseen speakers and out-of-distribution conditions.
 
 This setup yields $316$ recordings per room, for each microphone distance (close or far), and for each noise condition: Babble, Music, Television, and None (i.e., no background noise). Further details about the VOiCES devkit dataset can be found in \cite{richey2018voices}.

\begin{table}[!h]
\caption{\it Comparison of VAD performance across different feature-model combinations (AUC) using small-scale training and test sets on the LibriVAD-NonConcat dataset.}
   \centering
    \resizebox{\textwidth}{!}{
        \begin{tabular}{|l|r|c|c|c|c|c|c|c|c|c|c|} \hline 
         \textbf{Model } & \textbf{SNR} &  \multicolumn{6}{|c|}{\textbf{Seen}}  & \multicolumn{3}{|c|}{\textbf{Unseen}}   &\textbf{Average} \\
         &   & \textbf{Nature } & \textbf{Office } & \textbf{Public } & \textbf{Street } & \textbf{Transport} & \textbf{City } & \textbf{Babble} & \textbf{SSN} & \textbf{Domestic } &  \\ \hline \hline
         \multirow{6}{*}{raw-CLDNN} 
         & -5 & .8229    & .8501        & .7104  & .8122         & .9049         & .7402         & .5254        & .6635 & .773   &  \\
         & 0 & .8738     & .8947        & .7983  & .872      & .931      & .8164         & .6008            & .7719     & .8265  &  \\
         & 5 & .9079     & .9218        & .8624  & .9094         & .9481         & .8683         & .6969        & .8479     & .8718      &  \\
         & 10 & .9306    & .9386        & .9019  & .9345         & .9577         & .9044         & .7913        & .896      & .9105 &  \\
         & 15 & .9479    & .9505        & .9281  & .9502         & .9631         & .9309        & .866      & .9257         & .939       &  \\
         & 20 & .9562    & .9581        & .9451  & .959      & .966          & .9476    & .9137 & .9432 & .9553  &  \\
         & Avg & .9065    & .9189    & .8577  & .9062     & .9451     & .8679 & .7323  & .8413     & .8793 & .8728 \\ \hline
         \multirow{6}{*}{MFCC-bDNN}
         & -5 & .8419    & .8941        & .7233  & .8245         & .9188         & .7951         & .4578         & .7413        & .9088 &  \\
         & 0  & .8891    & .9213        & .8121  & .8778         & .9399         & .8591         & .4915         & .8364        & .9357 &  \\
         & 5  & .9173    & .9388        & .88    & .9134         & .9521         & .9029         & .5541         & .8992        & .9513 &  \\
         & 10 & .9333    & .9474        & .921   & .9351         & .9577         & .9292         & .6244         & .9352        & .9592 &  \\
         & 15 & .9418    & .9507        & .943   & .9468         & .9593         & .9439         & .69       & .9543    & .9627 &  \\
         & 20 & .9466    & .9518        & .955   & .9524         & .9592         & .9519         & .7473         & .9636        & .9639 &  \\
         & Avg & .9116  & .9340   & .8724  & .9083    & .9478     & .8970 & .5941   & .8883    & .9469 & .8785 \\ \hline 
         \multirow{6}{*}{GFCC-bDNN} 
         & -5 & .8283    & .8869        & .6996  & .8002   & .9025       & .7742         & .4647         & .7189        & .9021 &  \\
         & 0  & .8691    & .9062        & .7855  & .85     & .9201       & .835  & .4878         & .8199        & .9268 &  \\
         & 5  & .8943    & .9174        & .8503  & .8837   & .931        & .8765         & .5339         & .8853        & .9402  &  \\
         & 10 & .9081    & .9208        & .8898  & .9021   & .9346       & .9001         & .5824     & .9218    & .9461  &  \\
         & 15 & .9142    & .9188        & .9096  & .9091   & .9328       & .9108         & .6238         & .9402        & .9467  &  \\
         & 20 & .9147    & .9138        & .9182  & .9112   & .9286       & .9132         & .6556         & .9469        & .9433  &  \\
         & Avg & .8881     & .9106    & .8421  & .8760   & .9249   & .8683 & .5580    & .8721    & .9342 & .8527 \\ \hline
         \multirow{6}{*}{Silero-VAD}
         &-5 & .6647&.8689& .6305& .649 & .9391& .7129& .5065& .7372& .9041& \\
         & 0 &.7563& .9046& .7259& .7899& .9581& .7245& .5831& .8055& .9462&  \\
         & 5 &.9595& .928& .742  & .8747& .9648& .804 & .7428& .7895& .9593&   \\
         & 10&.9688& .9494& .8416& .9644& .9696& .9573& .8924& .8237& .9677& \\
         & 15&.9735& .9633& .9639& .9707& .9742& .9673& .9417& .8835& .9727&  \\
         & 20&.9764& .9711& .9709& .9747& .9767& .9733& .9662& .972 & .9755&  \\
         & Avg&.8832& .9308&.8124& .8705& .9637& .8565& .7721& .8352& .9542&  .8754\\ \hline 
         \multirow{6}{*}{MFCC-ViT} 
         & -5 & .9552   & .9434  & .8694         & .948 & .9695  & .9094        & .6123  & .8937        & .9689  &  \\
         & 0 & .9741    & .9647  & .9427         & .969     & .9783      & .9495         & .7475        & .9495     & .978       &  \\
         & 5 & .9815    & .9762  & .9668         & .9783        & .9831  & .972  & .8662            & .9712         & .984       &  \\
         & 10 & .9849   & .9819  & .9774         & .9844        & .986   & .9801        & .9275 & .9788     & .9857      &  \\
         & 15 & .9876   & .9862  & .9814         & .987     & .9878      & .9849          & .957            & .9823         & .988       &  \\
         & 20 & .9888   & .9864  & .9849         & .9881        & .9875  & .9867         & .9708        & .9848     & .989       &  \\
         & Avg & .9787    & .9731  & .9538     & .9758    & .982   & .9638 & .8469   & .9601     & .9823  &   {\bf .9574}\\ \hline 
         \multirow{6}{*}{GFCC-ViT} 
         & -5 & .9545   & .9475  & .8793        & .9422  & .9736         & .9044         & .6355         & .9167         & .9699         &  \\
         & 0  & .9712   & .9695  & .9444        & .9702  & .9784         & .9526         & .7768         & .9583         & .9798         &  \\
         & 5  & .9784   & .9779  & .9676        & .9802  & .9835         & .9734         & .9041         & .9724         & .9851         &  \\
         & 10 & .9837   & .9834  & .9768        & .9849  & .9865         & .9806         & .9505         & .9786         & .9877         &  \\
         & 15 & .9868   & .9858  & .9825        & .9877  & .9875         & .9838         & .9697         & .9847         & .9875         &  \\
         & 20 & .989        & .9872      & .9859        & .988   & .9887         & .9862         & .9783         & .9872         & .9883         &  \\
         & Avg& .9773    & .9752  & .9561    & .9755  & .983      & .9635     & .8692     & .9663     & .9831     &  \textbf{.9610} \\ \hline 

    \end{tabular}
    }
     \label{tab:LibriSpeech-NonConcat-small}
     \end{table}
     

\begin{table}[!h]
\caption{\it Comparison of VAD performance across different feature-model combinations (AUC) using small-scale training and test sets on the LibriVAD-Concat dataset.}
    \centering
    \resizebox{\textwidth}{!}{
    \begin{tabular}{|l|r|c|c|c|c|c|c|c|c|c|c|} \hline
    \textbf{Model } & \textbf{SNR} &  \multicolumn{6}{|c|}{\textbf{Seen}}  & \multicolumn{3}{|c|}{\textbf{Unseen}}   &\textbf{Average} \\
    &      & \textbf{Nature} & \textbf{Office} & \textbf{Public} & \textbf{Street } & \textbf{Transport} & \textbf{City } & \textbf{Babble} & \textbf{SSN} & \textbf{Domestic} &  \\ \hline \hline 
     \multirow{6}{*}{raw-CLDNN} 
     &-5&.906& .9068& .7253& .7889& .9156& .7693& .524& .6815& .5259& \\
&0&.9377& .939& .8239& .8705& .9437& .8482& .5996& .7946& .6586& \\
&5&.9566& .9558& .8896& .915& .961& .899& .7011& .8696& .7826& \\
&10&.9686& .9675& .9267& .9403& .9721& .9282& .7986& .9103& .8785& \\
&15&.9762& .9752& .9479& .9548& .9785& .9468& .8742& .9367& .9332& \\
&20&.9815& .9801& .9599& .9633& .9824& .9578& .9221& .9528& .9547& \\ 
&Avg&.9544& .9540& .8788& .9054& .9588& .8915& .7366& .8575& .7889& .8807\\ \hline 
    \multirow{6}{*}{MFCC-bDNN}
    &-5&.9522& .9621& .7432& .7556& .9391& .7902& .4297& .7604& .953& \\
&0 &.9712& .9757& .8492& .85& .9594& .8816& .4681& .8719& .9695& \\
&5&.9803& .9823& .9179& .9131& .9706& .9373& .5426& .9342& .978& \\
&10&.984& .9851& .9523& .9472& .9762& .9635& .6241& .9649& .9821& \\
&15&.9854& .9863& .9692& .9648& .9789& .9757& .6936& .9791& .984& \\
&20&.9858& .9865& .9768& .9727& .9801& .9814& .745& .9843& .9845& \\
&Avg&.9764& .9796& .9014& .9005& .9673& .9216& .5838& .9158& .9751& .9026\\ \hline
    \multirow{6}{*}{GFCC-bDNN}
    &-5&.9468& .9538& .717& .7185& .9167& .7746& .4344& .7527& .9492& \\
&0&.9635& .9685& .8174& .8104& .9361& .8621& .464& .864& .9642& \\
&5&.9721& .9754& .8852& .8746& .9483& .9181& .5214& .9274& .973& \\
&10&.9758& .9779& .9215& .9101& .9553& .9457& .5806& .9591& .977& \\ 
&15&.9769& .9781& .9402& .9283& .9595& .9584& .6264& .9737& .9779& \\
&20&.9769& .9772& .9482& .936& .9617& .9635& .6564& .9789& .9768& \\
&Avg&.9686& .9718& .8715& .8629& .9462& .9037& .5472& .9093& .9696& .8834\\ \hline

    \multirow{6}{*}{SileroVAD}
   &-5&.796& .9665& .6654& .681& .9698& .7564& .5211& .6989& .81& \\
&0&.9158& .9786& .8575& .7882& .9798& .812& .6574& .8322& .8718& \\
&5&.9861& .9836& .8569& .8525& .9842& .8738& .818& .8862& .9801& \\
&10&.9871& .9865& .9594& .9727& .9869& .9697& .9024& .8847& .9829& \\
&15&.988& .9879& .9803& .9826& .9879& .9807& .9408& .94& .9842& \\ 
&20&.989& .9886& .9863& .9865& .9885& .9854& .9713& .9874& .9862& \\ 
&Avg&.9436& .9819& .8843& .8772& .9828& .8963& .8018& .8715& .9358& .9084\\ \hline 

\multirow{6}{*}{MFCC-ViT}
&-5&.9909& .9899& .8982& .9411& .9896& .9032& .6111& .9265& .9828& \\
&0&.9938& .993& .9672& .979& .9931& .963& .7745& .9763& .9885& \\
&5&.9948& .9947& .9867& .9893& .9951& .9858& .9135& .9884& .9922& \\
&10&.9958& .9954& .9922& .9927& .996& .9918& .9588& .9918& .9948& \\
&15&.9962& .9957& .9943& .9949& .9962& .9941& .9804& .9934& .9957& \\
&20&.9965& .9959& .9949& .9959& .9966& .9951& .987& .9943& .9962& \\
&Avg&.9946& .9941& .9722& .9821& .9944& .9721& .8708& .9784& .9917& {\bf .9723} \\  \hline
\multirow{6}{*}{GFCC-ViT}
&-5&.9906& .9908& .9112& .9478& .9885& .918& .6306& .9401& .9827& \\
&0&.9934& .9945& .970& .9817& .9933& .9713& .7758& .9798& .9896& \\
&5&.9945& .9959& .9854& .9911& .9952& .9862& .8989& .9905& .9929& \\
&10&.9958& .9965& .9916& .9936& .996& .9917& .9548& .9933& .9949& \\
&15&.9964& .9965& .9941& .9949& .9964& .9939& .978& .9949& .9958& \\
&20&.9964& .9969& .9953& .9955& .9963& .9951& .9857& .9953& .9963& \\
&Avg&.9945& .9951& .9746& .9841& .9942& .9760& .8706& .9823& .9920& {\bf .9737} \\ \hline
\end{tabular}}
        \label{tab:LibriSpeech-Concat-small}
\end{table}

\section{Results and Discussions}\label{Results}
In this section,  we first analyze and compare different feature representations and modeling techniques using the proposed LibriVAD-NonConcat and LibriVAD-Concat datasets under the small-scale setup. We then further study the best-performing feature-model combination  by training on  datasets of varying sizes to evaluate its generalization capability. 

\subsection{Analysis of VAD Performance on LibriVAD-NonConcat  and LibriVAD-Concat}
Tables  \ref{tab:LibriSpeech-NonConcat-small} and \ref{tab:LibriSpeech-Concat-small}  compare VAD performance across different feature-model combinations using the LibriVAD-NonConcat and LibriVAD-Concat datasets under the small-scale setup.
Table \ref{tab:LibriSpeech-NonConcat-small} shows that ViT-based systems, with either MFCC or GFCC features, consistently achieve the highest AUC values, confirming that ViT is the most effective architecture for the LibriVAD-NonConcat dataset. Between the two feature types, GFCC performs slightly better than MFCC. Among the various noisy conditions, Babble noise yields the lowest average AUC scores across SNR levels, likely because its similarity to human speech makes distinguishing speech from non-speech more challenging.  Table \ref{tab:LibriSpeech-Concat-small} presents the results for the LibriVAD-Concat dataset, where ViT again achieves the highest AUC values for both MFCC and GFCC, with GFCC slightly outperforming MFCC. 

Overall, the ViT-based systems trained on the LibriVAD-Concat variant consistently deliver higher average AUC scores across all feature types and models compared to those trained on the LibriVAD-NonConcat variant.

\begin{table}[h]
\centering
\caption{Effect of varying number of encoder layers in the ViT compared to bDNN and CLDNN for MFCC based system on the LibriVAD-NonConcat and LibriVAD-Concat datasets. All evaluations are performed under the respective small-scale training and test conditions, and the average AUC across all noise conditions is reported.} 
\footnotesize{
\begin{tabular}{|l|c|c|c|c|c|c|c|c|c|c|}\hline 
                    & raw-  & bDNN    & \multicolumn{8}{|c|}{ \#of encoder layers in ViT}\\
                    &  CLDNN &         &      1        & 2        & 4         &8         & 10    & 12      & 14   & 16 \\ \hline 
Model Parameters    & 157k   & 406.5k  & 471.9k        & 916.2k   & 1.81M     & 3.58M    &4.47M  & 5.36M   &6.25M & 7.14M   \\
LibriVAD-NonConcat  & .8728  & .8785   &.9565         & .9573   & .9635    &.9574     & .9636 &  .9574  & .9621 & .9588  \\ 
LibriVAD-Concat     & .8807  & .9026   & .9697         & .9710      & .9732  &.9733     & .9725 &  .9723 &.9693  &.9735\\ \hline 
\end{tabular}
}
\label{table:no_end_lyer}
\end{table}

We further analyze the effect of the number of encoder layers in the ViT-MFCC system, as shown in  Table \ref{table:no_end_lyer}, for both the Librivad-NonConcat and Librivad-Concat datasets under small-scale training and test conditions. Table \ref{table:no_end_lyer} reports the average AUC across all evaluation noise conditions.  The results show only marginal changes when deviating from the default $12$‑layer configuration, with a slight improvement at $4$ layers and no degradation even at $2$ layers, indicating that model depth alone is not the primary factor driving the performance gains of ViT‑MFCC compared with other architectures. It is worth noting that the 2‑ and 4‑layer ViT models not only achieve performance comparable to the default 12‑layer system on the in‑domain test set, but also maintain a similar level of generalization on the VOiCES dataset is weaker.\footnote{Table \ref{table:MFA-voices}(b) reports the VOiCES results for the default 12‑layer ViT model trained on LibriVAD-Concat. For comparison, the 4‑layer variant achieves AUC scores of 0.9409 (clo) and  0.8719 (far), whereas the 12‑layer model obtains 0.9358 (clo) and 0.8588 (far).}

\begin{figure}[!h]
  \centering
  \subfloat[LibriVAD-NonConcat]{\includegraphics[width=16.0cm, height=6.0cm]{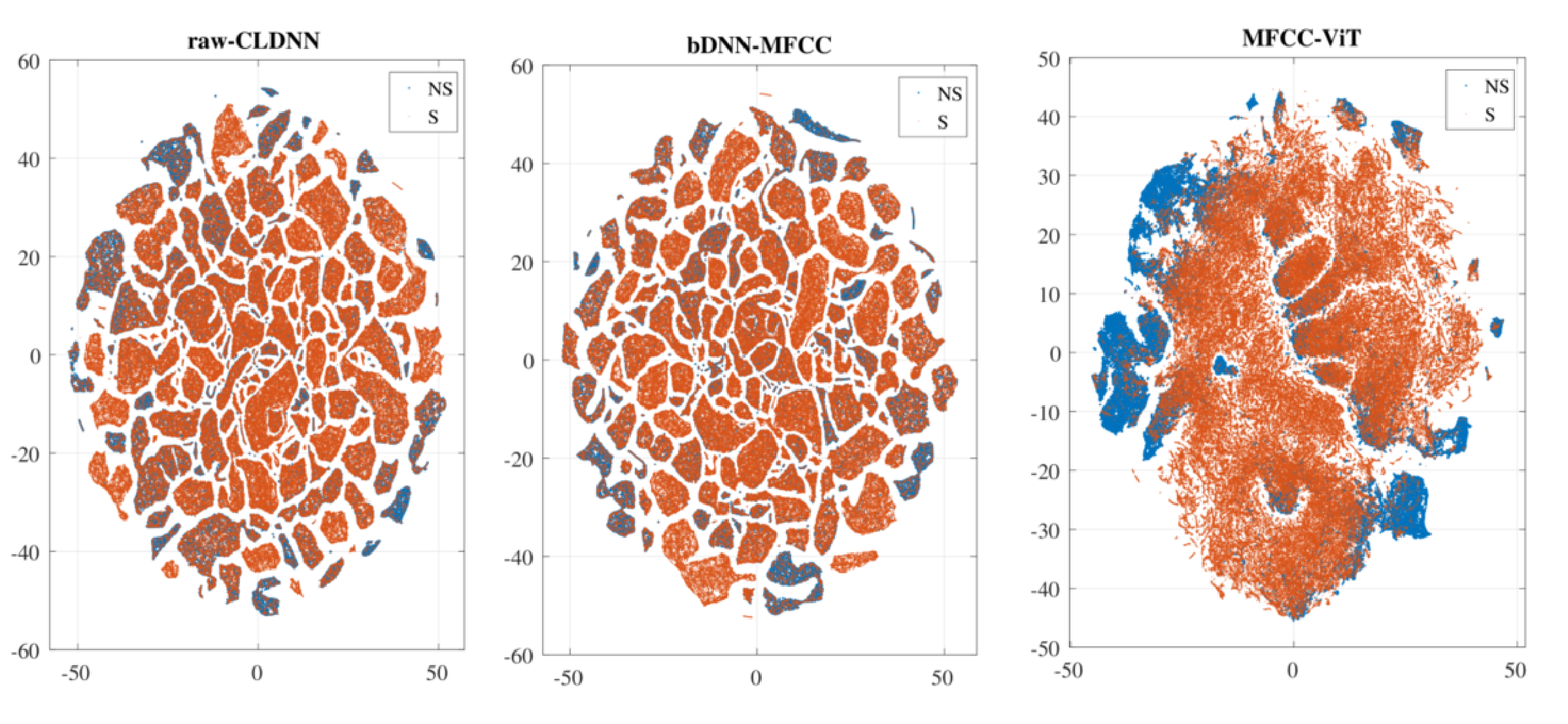}}
  \hspace{-0.5cm}%
  \subfloat[LibriVAD-Concat.] {\includegraphics[width=16.0cm, height=6.0cm]{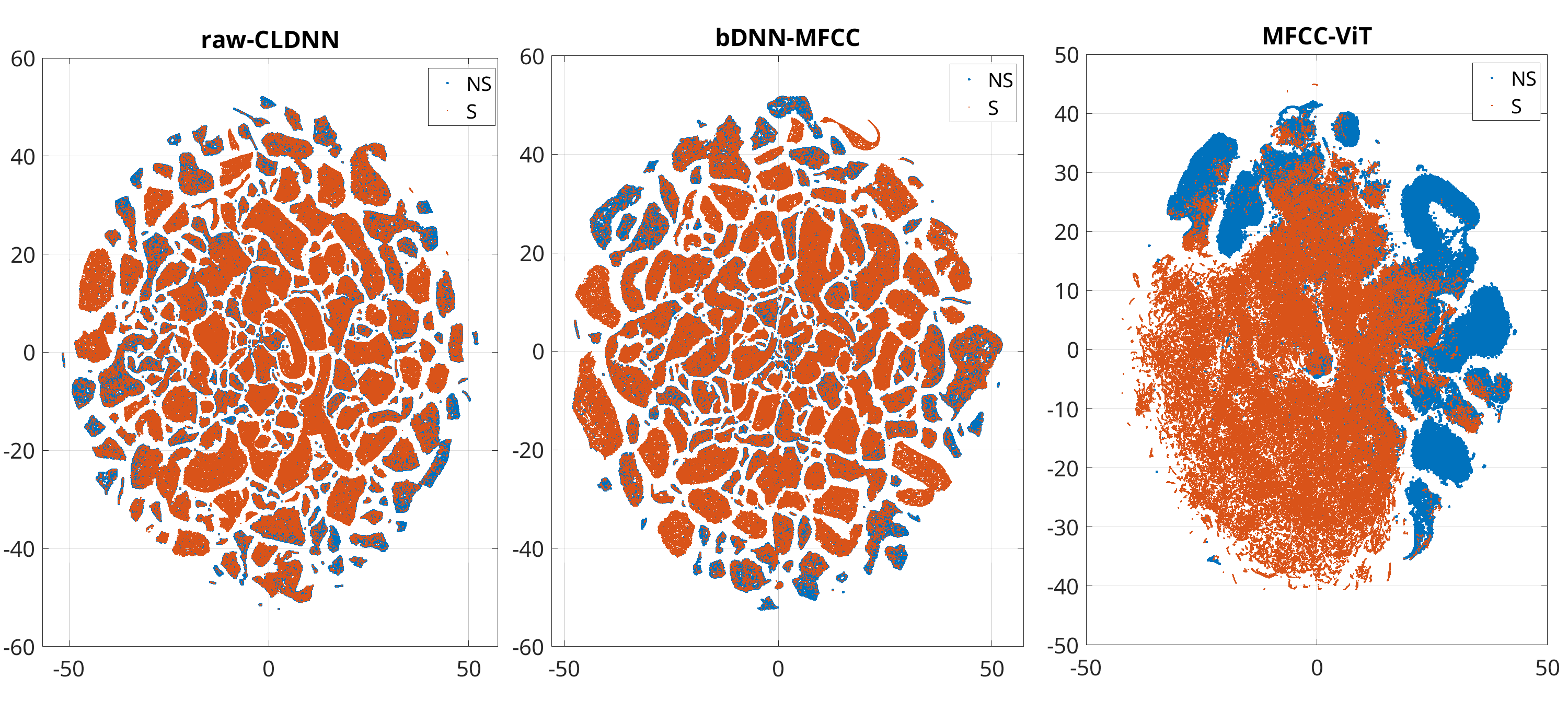}} 
    \caption{Illustration of the t-SNE \cite{tsne} scatter plots of the unseen evaluation noisy data based on the last layer hidden  representations in the respective systems for the LibriVAD-NonConcat and LibriVAD-Concat datasets under the small-scale condition. Each system uses the same data points for the respective datasets. $NS$ (blue) and $S$ (orange) indicate the non-speech and speech classes, respectively. (a)  LibriVAD-NonConcat. (b)  LibriVAD-Concat.}
    \label{fig:t_sne_plot}
\end{figure}

To further  investigate why the ViT-MFCC is superior to the bDNN and CLDNN, we scatter-plot the last-layer representations of each respective system for the unseen evaluation data (this dataset is more challenging due to its unseen nature) using the T-SNE toolkit \cite{tsne} for the Librivad-NonConcat and Librivad-Concat datasets. As illustrated in Fig. \ref{fig:t_sne_plot}, the ViT-MFCC system displays greater discriminative capability between speech and non-speech compared to bDNN and CLDNN.   

\begin{table}[h]
\centering
\caption{Comparison of model parameters, training time, and inference time for the different systems on the LibriVAD-NonConcat and LibriVAD-Concat datasets under the small-scale training and test conditions. }
\footnotesize{
\begin{tabular}{|l|c|c|c|c|c|}\hline 
Model                                         &  Parameters  & \multicolumn{2}{|c|}{ Training time (Seconds per Epoch)}     & \multicolumn{2}{|c|}{ Evaluation time (Seconds/RTF)}\\
                                              &              & LibriVAD-NonConcat   &  LibriVAD-Concat  & LibriVAD-NonConcat   &  LibriVAD-Concat\\ \hline 
raw-CLDNN                                     &  157k        & 305.48               &  383.23           &   9.58/8.58$\times 10^{-4}$              & 11.96/8.10$\times 10^{-4}$             \\         
MFCC-bDNN                                     &  406.5k      & 93.03                &  125.34           &  2.92/2.62$\times 10^{-4}$               &  3.32/2.24$\times 10^{-4}$             \\
MFCC-ViT:                                     &              &                      &                   &                      &                   \\    
 - 1 Encoder layer          & 471.9k            & 18.90  &  21.68     & 6.64/5.94$\times 10^{-4}$     & 6.75/4.57$\times 10^{-4}$       \\  
- 12 Encoder layers                           &  5.36M       & 147.68               &  190.55           &  36.47/32.67$\times 10^{-4}$               &  57.29/38.81$\times 10^{-4}$           \\ \hline

\end{tabular}
}
\label{table:para_train_test_time}
\end{table}

Table \ref{table:para_train_test_time} compares model sizes, training time per epoch, and inference time of different systems on the LibriVAD-NonConcat and LibriVAD-Concat datasets under the small-scale training and test sets, as described in Table \ref{table:libriVAD_summary}. All experiments were conducted on a computer equipped with an Intel(R) 12th Gen Core(TM) i9-12900K, 128GB RAM, a 4TB SSD, an NVIDIA GeForce RTX3060 GPU (12 GB), and Ubuntu 24.04.3 LTS operating system.  

For the bDNN and ViT-based systems, precomputed MFCC features are used during both training and inference. Consequently, feature extraction is excluded from the reported training and inference time measurements. In contrast, raw-CLDNN processes raw audio waveforms directly. 
To isolate and compare the computational cost of the neural network architectures, all timing measurements are conducted using prepared inputs, either pre-segmented audio signals or precomputed features. Therefore, the reported timing results reflect the computational efficiency of the neural network models themselves rather than differences in preprocessing implementations. 
The Real-Time Factor (RTF) metric is defined as the ratio between the total inference time and the duration of the evaluated audio data.  As shown in Table \ref{table:para_train_test_time}, the bDNN model requires less inference time than to the ViT-based models, likely due to the differences in model architectures and data-processing pipelines. However, the ViT model with a single encoder layer has a parameter count comparable to that of the bDNN while requiring less training time. In addition, the single-layer ViT achieves absolute AUC improvement of  approximately 7.80 and 6.71 percentage points over the bDNN on the LibriVAD-Nonconcat and LibriVAD-Concat datasets, respectively.

\subsection{Effect of Training Data Size on Generalization}
In this section, we examine how the size of training data influences VAD performance. 
The performance of the ViT-MFCC and ViT-GFCC are comparable  as shown in Tables \ref{tab:LibriSpeech-NonConcat-small} and \ref{tab:LibriSpeech-Concat-small}, and we focus on the ViT model with MFCC.  

Table \ref{table:small-medium-large-data-size-LibriSpeech} shows the impact of training data size on VAD performance, 
with all evaluations conducted on the small test set to ensure consistency across models, regardless of whether they were trained on small, medium, or large datasets. 

The results show that increasing the training data size from small to medium improves AUC, whereas further scaling to the large dataset leads to a slight decline in performance. This finding contrasts with the common expectation that more training data consistently improves model generalization. 
Nevertheless, the observed performance differences are relatively small, and the present experiments do not allow us to conclusively attribute the decline to increased data size and diversity alone. Alternative factors, including suboptimal training schedules, insufficient model capacity, or the need for additional hyperparameter tuning when training on larger datasets, may also contribute. 


Table \ref{table:cleantest} presents AUC scores for clean test data evaluation, comparing systems trained with different data sizes using the ViT-MFCC model.  Consistent with Table \ref{tab:LibriSpeech-Concat-small},  the LibriVAD-Concat dataset combined with ViT-MFCC achieves the highest average AUC. Overall, the results indicate that all VAD systems perform comparably on clean test data.
 
\begin{table}[!h]
\caption{\it Impact of training data size on VAD performance (AUC) using the ViT model with MFCC features on the LibriVAD-NonConcat and LibriVAD-Concat datasets. All evaluations are conducted on the respective small test sets to ensure consistency.}
\begin{subtable}{0.99\textwidth}
\caption{Results with LibriVAD-NonConcat}
 \centering\
    \resizebox{\textwidth}{!}{
    \begin{tabular}{|l|r|c|c|c|c|c|c|c|c|c|c|}\hline
         \textbf{System [Dataset] } & \textbf{SNR} & \multicolumn{6}{|c|}{\textbf{Seen}}  & \multicolumn{3}{|c|}{\textbf{Unseen}}   &\textbf{Average} \\
        &   &\textbf{Nature } & \textbf{Office } & \textbf{Public } & \textbf{Street } & \textbf{Transport} & \textbf{City } & \textbf{Babble } & \textbf{SSN } & \textbf{Domestic } &  \\\hline \hline 
         \multirow{6}{*}{MFCC-ViT [Small]} 
         & -5 & .9552   & .9434  & .8694         & .948     & .9695      & .9094        & .6123  & .8937        & .9689  &  \\
         & 0  & .9741   & .9647  & .9427         & .969     & .9783      & .9495         & .7475        & .9495     & .978       &  \\
         & 5 & .9815    & .9762  & .9668         & .9783        & .9831  & .972  & .8662            & .9712         & .984       &  \\
         & 10 & .9849   & .9819  & .9774         & .9844        & .986   & .9801        & .9275 & .9788     & .9857      &  \\
         & 15 & .9876   & .9862  & .9814         & .987     & .9878      & .9849          & .957            & .9823         & .988       &  \\
         & 20 & .9888   & .9864  & .9849         & .9881        & .9875  & .9867         & .9708        & .9848     & .989       &  \\
         & Avg & .9787    & .9731  & .9538     & .9758    & .982   & .9638 & .8469   & .9601     & .9823  &   .9574\\ \hline
         \multirow{6}{*}{MFCC-ViT [Medium]}
         & -5 & .9756    & .9712          & .9015        & .9582         & .9836         & .9248         & .6504 & .9454 & .9782 &  \\
         & 0  & .9871    & .9845          & .9653        & .9819         & .9903         & .9708         & .8036 & .9763 & .9872 &  \\
         & 5  & .9915    & .9902          & .9839        & .9898         & .9932         & .9855         & .915  & .986  & .9913 &  \\
         & 10 & .9936    & .993       & .9902    & .993      & .9945     & .9911         & .9628 & .9904 & .9933 &  \\
         & 15 & .9946    & .9943          & .9929        & .9944         & .9951         & .9934         & .9806 & .9927 & .9944 &  \\
         & 20 & .9951    & .9949          & .9943        & .9951         & .9954         & .9946         & .9881 & .994  & .995  &  \\
         & Avg & .9895     & .9880      & .97135   & .9854     & .9920     & .9767     & .8834& .9808 & .9899 &  {\bf .9730}\\ \hline 
         \multirow{6}{*}{MFCC-ViT [Large]} 
         & -5 & .9766    & .9694 & .8939         & .9582 & .9834         & .9265         & .6254         & .9414         & .9771  &  \\
         & 0  & .9879    & .9835 & .9621         & .982  & .9904         & .9717         & .7764         & .9786         & .987  & \\
         & 5  & .9921    & .9901 & .9838         & .9902 & .9935         & .9866         & .8976         & .9882         & .9915  &   \\
         & 10 & .994     & .9931 & .9906         & .9933 & .9949         & .9918         & .9553         & .9917         & .9936  &  \\
         & 15 & .9949    & .9946 & .9933         & .9947 & .9955         & .9939         & .9776         & .9934         & .9947  &  \\
         & 20 & .9955    & .9952 & .9946         & .9954 & .9958         & .995      & .9864     & .9945         & .9953  &  \\
         & Avg& .9901    & .9876 & .9697     & .9856 & .9922     & .9775     & .8697     & .9813      & .9898 &  .9715 \\ \hline 
 
        \end{tabular}}
        \label{table:LibriSpeech-NonConcat-small-medium-large}
        \end{subtable}\\
    \begin{subtable}{0.99\textwidth}
    \caption{Results with LibriVAD-Concat}    
    \centering
    \resizebox{\textwidth}{!}{
    \begin{tabular}{|l|r|c|c|c|c|c|c|c|c|c|c|} \hline 
              \textbf{system [Dataset] } & \textbf{SNR} & \multicolumn{6}{|c|}{\textbf{Seen}}  & \multicolumn{3}{|c|}{\textbf{Unseen}}   &\textbf{Average} \\
         &                 & \textbf{Nature } & \textbf{Office } & \textbf{Public} & \textbf{Street} & \textbf{Transport} & \textbf{City} & \textbf{Babble} & \textbf{SSN } & \textbf{Domestic}  & \\ \hline \hline 
         \multirow{6}{*}{MFCC-ViT [Small]} 
         &-5&.9909& .9899& .8982& .9411& .9896& .9032& .6111& .9265& .9828& \\
&0&.9938& .993& .9672& .979& .9931& .963& .7745& .9763& .9885& \\
&5&.9948& .9947& .9867& .9893& .9951& .9858& .9135& .9884& .9922& \\
&10&.9958& .9954& .9922& .9927& .996& .9918& .9588& .9918& .9948& \\
&15&.9962& .9957& .9943& .9949& .9962& .9941& .9804& .9934& .9957& \\
&20&.9965& .9959& .9949& .9959& .9966& .9951& .987& .9943& .9962& \\
&Avg&.9946& .9941& .9722& .9821& .9944& .9721& .8708& .9784& .9917& .9723   \\ \hline 
         \multirow{6}{*}{MFCC-ViT [Medium]}
&-5&.9952& .9938& .905& .9429& .9934& .9408& .6357& .9641& .9851& \\
&0&.9966& .9961& .9735& .9845& .9964& .9822& .8036& .9895& .9915& \\
&5&.9974& .9971& .9915& .9939& .9976& .9926& .9243& .9937& .9946& \\
&10&.9979& .9976& .9957& .9965& .9981& .9953& .9706& .996& .9966& \\
&15&.9983& .998& .997& .9973& .9983& .9967& .9886& .997& .9978& \\
&20&.9984& .9982& .9977& .9979& .9984& .9975& .9944& .9976& .9982& \\
&Avg&.9973& .9968& .9767& .9855& .9970& .9841& .8862& .9896& .9939& {\bf .9785} \\ \hline
          \multirow{6}{*}{MFCC-ViT [Large]}
&-5&.9944& .9935& .8762& .9392& .9919& .9131& .6005& .9621& .9855& \\
&0&.9964& .9961& .9682& .986& .9959& .9728& .7698& .9882& .9914& \\
&5&.9975& .9972& .9899& .9946& .9975& .9902& .9196& .9935& .995& \\
&10&.998& .9977& .9952& .9964& .9978& .9948& .971& .9957& .9967& \\
&15&.9983& .9981& .9969& .9973& .9982& .9965& .9867& .9969& .9976& \\
&20&.9983& .998& .9976& .9977& .9981& .9974& .9924& .9975& .9979& \\
&Avg&.9971& .9967& .9706& .9852& .9965& .9774& .8733& .9889& .9940& .9755\\ \hline
\end{tabular}}
        \label{table:LibriSpeech-Concat-small-medium-large}
        \end{subtable}
        \label{table:small-medium-large-data-size-LibriSpeech}
\end{table}

\begin{table}[!h]
\caption{\it Comparison  of VAD performance (AUC) across different training data sizes and datasets on clean test data under the small-scale evaluation setup using the ViT-MFCC model.}
\centering
\footnotesize{
\begin{tabular}{|l|ll|l|}\cline{1-4} 
Train data    & Train                  & Test                   & AUC  \\        
 Size        & Database               & Database                & \\ \hline  \hline 
Small  &LibriVAD-NonConcat   &LibriVAD-NonConcat   & .9865  \\ 
Small  &LibriVAD-Concat      &LibriVAD-Concat      & .9934 \\  \hline 
Medium &LibriVAD-NonConcat   & LibriVAD-NonConcat  &  .9947      \\
       &LibriVAD-Concat      &LibriVAD-Concat      & .9977   \\ \hline 
Large  &LibriVAD-NonConcat   & LibriVAD-NonConcat  &  .9951     \\
       &LibriVAD-Concat      &LibriVAD-Concat      &  .9962 \\      \hline 
\end{tabular}
}
\label{table:cleantest}
\end{table}

\begin{table*}[!pt]
\caption{\it Effect of training-test data mismatch on VAD performance (AUC) using the MFCC-ViT technique on the LibriVAD-NonConcat and LibriVAD-Concat datasets. Evaluations are conducted under the small-scale setup across models trained with small, medium, and large data sizes within each respective dataset.}
\begin{subtable}{0.99\textwidth}
\caption{Model trained on LibriVAD-NonConcat and evaluated on LibriVAD-Concat}
 \centering
    \resizebox{\textwidth}{!}{
    \begin{tabular}{|l|r|c|c|c|c|c|c|c|c|c|c|} \hline 
     \textbf{Train data } & \textbf{SNR} &  \multicolumn{6}{|c|}{\textbf{Seen}}  & \multicolumn{3}{|c|}{\textbf{Unseen}}   &\textbf{Average} \\
     Size &    & \textbf{Nature } & \textbf{Office } & \textbf{Public } & \textbf{Street } & \textbf{Transport} & \textbf{City} & \textbf{Babble } & \textbf{SSN } & \textbf{Domestic } &  \\ \hline \hline 
     \multirow{6}{*}{Small}
        &-5 & .9683 & .9670 & .8774 & .9144 & .9756 & .8977 & .5980 & .8944 & .9655 & \\
        &0 & .9842 & .9791 & .9503 & .9642 & .9841 & .9579 & .6984 & .9663 & .9818 & \\
        &5 & .9897 & .9862 & .9765 & .9818 & .9891 & .9794 & .8195 & .9810 & .9879 & \\
        & 10 & .9918 & .9893 & .9866 & .9881 & .9915 & .9882 & .9007 & .9882 & .9918 & \\
        & 15 & .9930 & .9914 & .9892 & .9913 & .9920 & .9914 & .9469 & .9899 & .9925 & \\
        & 20 & .9932 & .9923 & .9913 & .9925 & .9927 & .9927 & .9645 & .9917 & .9933 & \\
        & Avg & .9867 & .9842 & .9619 & .9720 & .9875 & .9679 & .8213 & .9686 & .9855 & .9595\\ \hline
     \multirow{6}{*}{Medium}
        &-5 & .9865 & .9887 & .9184 & .9595 & .9918 & .9621 & .6247 & .9558 & .9841 & \\
        &0 & .9952 & .9945 & .9773 & .9851 & .9957 & .9887 & .7644 & .9889 & .9932 & \\
        &5 & .9975 & .9969 & .9918 & .9942 & .9974 & .9955 & .8903 & .9938 & .9959 & \\
        & 10 & .9983 & .9980 & .9954 & .9966 & .9980 & .9969 & .9576 & .9961 & .9971 & \\
        & 15 & .9985 & .9983 & .9970 & .9977 & .9983 & .9974 & .9811 & .9973 & .9980 & \\
        & 20 & .9985 & .9984 & .9979 & .9982 & .9986 & .9979 & .9910 & .9976 & .9983 & \\
        & Avg & .9958 & .9958 & .9796 & .9886 & .9966 & .9898 & .8682 & .9882 & .9944 & {\bf .9774}\\ \hline
     \multirow{6}{*}{Large}
        &-5 & .9832 & .9808 & .8952 & .9511 & .9873 & .9476 & .5985 & .9264 & .9789 & \\
        &0 & .9929 & .9901 & .9651 & .9802 & .9937 & .9838 & .7320 & .9818 & .9898 & \\
        &5 & .9970 & .9950 & .9872 & .9921 & .9965 & .9939 & .8600 & .9927 & .9946 & \\
        & 10 & .9981 & .9966 & .9938 & .9961 & .9976 & .9965 & .9396 & .9953 & .9967 & \\
        & 15 & .9983 & .9975 & .9964 & .9977 & .9979 & .9971 & .9707 & .9968 & .9972 & \\
        & 20 & .9982 & .9977 & .9974 & .9980 & .9980 & .9976 & .9839 & .9974 & .9978 & \\
        & Avg & .9946 & .9929 & .9725 & .9859 & .9952 & .9861 & .8474 & .9817 & .9925 & .9721\\ \hline
    \end{tabular}}\\[1.2ex]
    \caption{Model trained on LibriVAD-NonConcat and evaluated on LibriVAD-Concat}
 \centering
    \resizebox{\textwidth}{!}{
    \begin{tabular}{|l|r|c|c|c|c|c|c|c|c|c|c|} \hline 
     \textbf{Train data } & \textbf{SNR} &  \multicolumn{6}{|c|}{\textbf{Seen}}  & \multicolumn{3}{|c|}{\textbf{Unseen}}   &\textbf{Average} \\
     Size &    & \textbf{Nature } & \textbf{Office } & \textbf{Public } & \textbf{Street } & \textbf{Transport} & \textbf{City} & \textbf{Babble } & \textbf{SSN } & \textbf{Domestic } &  \\ \hline \hline 
     \multirow{6}{*}{Small}
        &-5 & .9661 & .9537 & .8315 & .9331 & .9740 & .8901 & .6137 & .9115 & .9656 & \\
        &0 & .9802 & .9720 & .9422 & .9682 & .9835 & .9600 & .7629 & .9640 & .9777 & \\
        &5 & .9854 & .9817 & .9726 & .9794 & .9869 & .9767 & .9029 & .9764 & .9844 & \\
        & 10 & .9878 & .9860 & .9841 & .9856 & .9892 & .9835 & .9511 & .9832 & .9884 & \\
        & 15 & .9890 & .9886 & .9882 & .9885 & .9903 & .9867 & .9697 & .9872 & .9895 & \\
        & 20 & .9903 & .9896 & .9889 & .9898 & .9911 & .9886 & .9796 & .9890 & .9902 & \\
        & Avg & .9831 & .9786 & .9512 & .9741 & .9858 & .9643 & .8633 & .9686 & .9826 & .9613\\ \hline
     \multirow{6}{*}{Medium}
        &-5 & .9721 & .9601 & .8616 & .9462 & .9833 & .9135 & .6185 & .9327 & .9747 & \\
        &0 & .9834 & .9784 & .9522 & .9710 & .9890 & .9684 & .7857 & .9727 & .9853 & \\
        &5 & .9882 & .9873 & .9780 & .9813 & .9914 & .9841 & .9044 & .9809 & .9895 & \\
        & 10 & .9905 & .9905 & .9868 & .9872 & .9924 & .9881 & .9561 & .9862 & .9921 & \\
        & 15 & .9923 & .9925 & .9895 & .9907 & .9935 & .9901 & .9759 & .9890 & .9932 & \\
        & 20 & .9934 & .9931 & .9919 & .9922 & .9941 & .9920 & .9836 & .9914 & .9937 & \\
        & Avg & .9867 & .9837 & .9600 & .9781 & .9906 & .9727 & .8707 & .9755 & .9881 & {\bf .9673}\\ \hline
     \multirow{6}{*}{Large}
        &-5 & .9660 & .9489 & .8215 & .9264 & .9774 & .8746 & .5514 & .9196 & .9704 & \\
        &0 & .9816 & .9747 & .9337 & .9692 & .9895 & .9603 & .7053 & .9685 & .9862 & \\
        &5 & .9873 & .9863 & .9745 & .9833 & .9921 & .9801 & .8599 & .9822 & .9904 & \\
        & 10 & .9908 & .9907 & .9852 & .9887 & .9929 & .9870 & .9423 & .9871 & .9925 & \\
        & 15 & .9934 & .9926 & .9900 & .9917 & .9940 & .9905 & .9732 & .9899 & .9935 & \\
        & 20 & .9935 & .9936 & .9927 & .9930 & .9942 & .9925 & .9841 & .9922 & .9938 & \\
        & Avg & .9854 & .9811 & .9496 & .9754 & .9900 & .9642 & .8360 & .9733 & .9878 & .9603\\ \hline
    \end{tabular}}
        \label{table:LibriSpeech-Concat-test-NonConcat-small}
        \end{subtable}
        \label{table:in-domain-cross-eval}
\end{table*}     

\begin{table}[!ht]
\caption{\it Comparison of VAD system performance based on EER and MinDCF under the small-scale evaluation setup.}
\begin{center}
\scriptsize{
\begin{tabular}{|l|ll|l|l|c|}\cline{1-6} 
Train data     & Train data                      & Test data                 & System    & EER (\%)  & MinDCF \\
  Size          & Concat /NonConcat                   &Concat /NonConcat               &           &           &         \\ \hline  \hline 
          & LibriVAD-NonConcat      &LibriVAD-NonConcat  &raw-CLDNN  & 19.29     & .0756  \\
Small     & \checkmark                 & \checkmark            &MFCC-bDNN  & 19.52     & .0919 \\
          & \checkmark                 & \checkmark            & GFCC-bDNN & 22.13     & .0953 \\
-         &  -                         &\checkmark             & Silero-VAD & 26.06     & .0593  \\        
          & \checkmark                 & \checkmark            & MFCC-ViT  & 10.40      & .0528  \\
          & \checkmark                 & \checkmark            & GFCC-ViT  & 9.30      & .0456  \\ 
          &                            &                       &           &           &       \\ 
          &LibriVAD-Concat          &LibriVAD-Concat     &raw-CLDNN        & 18.62     & .0725    \\ 
          & \checkmark& \checkmark                       &MFCC-bDNN        & 15.98     & .0774 \\
          & \checkmark& \checkmark                       & GFCC-bDNN       & 18.19     & .0829 \\
-         &  -        & \checkmark                       &Silero-VAD       & 27.08     & .0541 \\           
           & \checkmark& \checkmark                      & MFCC-ViT        & {\bf 7.19} & {\bf .0353} \\
           & \checkmark& \checkmark                      & GFCC-ViT        & {\bf 7.17 }  & .0380  \\ 
           &           &                                 &                 &          &         \\
           &LibriVAD-NonConcat      & LibriVAD-Concat   & MFCC-ViT  & 8.99      & .0453  \\        
           & LibriVAD-Concat        &LibriVAD-NonConcat  & MFCC-ViT  & 10.58    &  .0470    \\ \hline     
Medium     &LibriVAD-NonConcat      &LibriVAD-NonConcat  & MFCC-ViT  & 6.90      & .0348 \\ 
           &LibriVAD-Concat         &LibriVAD-Concat     &           & {\bf 5.92}  &  .0311 \\ \hline 
           &LibriVAD-NonConcat      &LibriVAD-NonConcat  & MFCC-ViT  & 7.10  & .0377   \\
Large      & LibriVAD-Concat        &LibriVAD-Concat     &           & 6.14  &  .0299\\ \hline 
 
\end{tabular}
}
\end{center}
\label{table:EER_DCF}
\end{table}

\subsection{Cross-Dataset Evaluation of VAD Systems}
Table \ref{table:in-domain-cross-eval} presents the cross-dataset evaluation results for the MFCC-ViT system on the LibriVAD-NonConcat and LibriVAD-Concat datasets. In this setup, the model is trained on one dataset and tested on the other, and vice versa. The AUC patterns observed here closely resemble those reported in Table \ref{table:small-medium-large-data-size-LibriSpeech}, where models were evaluated without cross-evaluation. 

The distinction between the LibriVAD-NonConcat and LibriVAD-Concat variants primarily affects the temporal structure of the training data and the continuity of the acoustic context presented to the model. In the LibriVAD-NonConcat setting, each utterance is treated as an independent sequence, and temporal context is limited to within-utterance dynamics. In contrast, the LibriVAD-Concat variant produces longer continuous audio streams by concatenating two utterances with inserted silence segments, resulting in extended temporal sequences.

From a modeling perspective, the LibriVAD-Concat formulation encourages learning temporal dependencies across utterance boundaries, including transitions between speech and silence over longer time scales. This setting more closely resembles real-world audio streams, where speech activity is continuous and utterance boundaries are not explicitly defined.  
By providing both variants, LibriVAD enables a controlled investigation of how temporal continuity and label structure affect model behavior, allowing researchers to choose the formulation that best aligns with their deployment needs.

\subsection{Performance of VAD Systems based on EER and MinDCF}
Table \ref{table:EER_DCF} presents the performance of different VAD systems trained with varying data sizes, evaluated using EER and MinDCF. While AUC is used as the primary metric due to its threshold-independent nature, EER and MinDCF provide complementary operating-point-based evaluations, particularly under varying speech-to-non-speech ratios. The proposed ViT-MFCC model consistently achieves the lowest EER and MinDCF values among all systems, whereas Silero-VAD exhibits the highest EER. 

Increasing the training data size, such as using the medium dataset, further reduces error rates, which aligns with the AUC trend observed in Table \ref{table:small-medium-large-data-size-LibriSpeech}. This finding reinforces the strong generalizability of the proposed ViT VAD system.

In the cross-dataset evaluation setting, training on the LibriVAD-Concat dataset and testing on the LibriVAD-NonConcat dataset leads to a slight increase in EER, consistent with the AUC pattern reported in Table \ref{table:in-domain-cross-eval}. {The consistency between AUC and EER trends further supports the robustness of the reported findings, despite the presence of class imbalance.}

\begin{figure}[h]
  \centering
  \subfloat[Train: Small LibriVAD-NonConcat subset; Test: Small LibriVAD-NonConcat subset.]{\includegraphics[width=.6\linewidth]{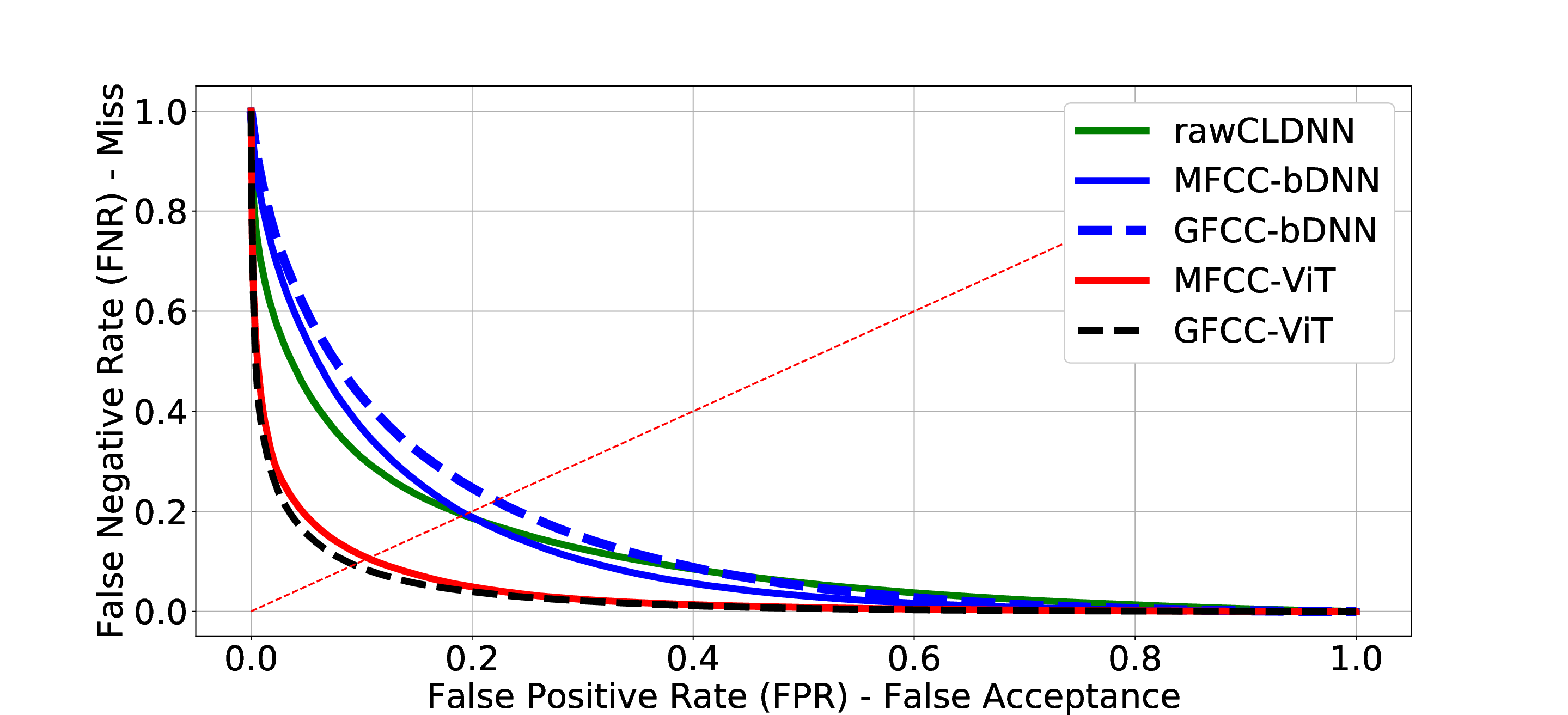}}
  \hspace{-0.5cm}%
  \subfloat[Train: Small LibriVAD-Concat subset; Test: Small LibriVAD-Concat subset.] {\includegraphics[width=.6\linewidth]{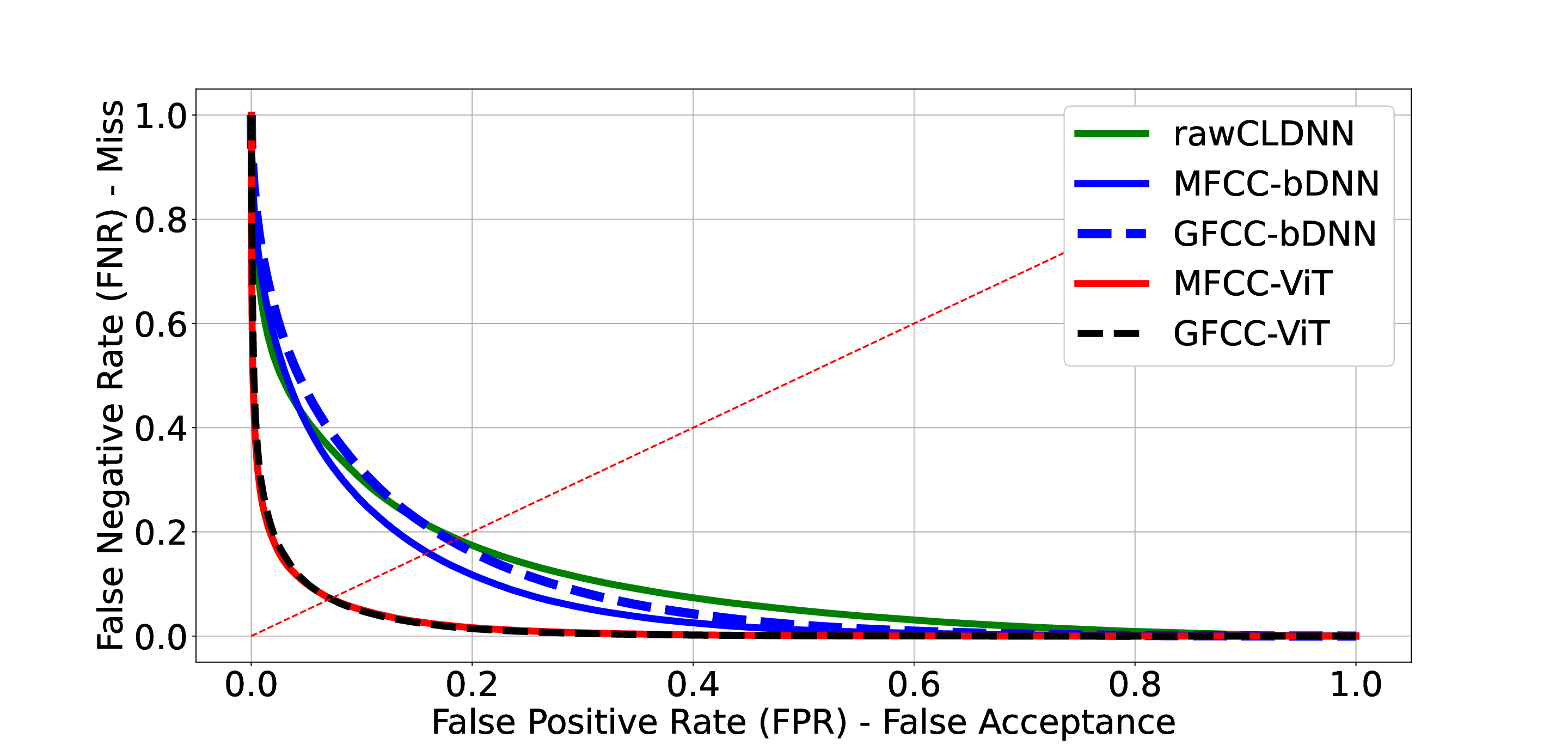}} 
    \caption{DET curves for different VAD models across a wide operating range, with scores aggregated over various noise types and SNR levels. (a) Model trained and evaluated on the small set of LibriVAD-NonConcat. (b) Model trained and evaluated on the small set of LibriVAD-Concat.}
   \label{fig:DET_small_small}
\end{figure}

Figure \ref{fig:DET_small_small} presents DET curves, plotting miss rate versus false acceptance rate across a wide operating range for the LibriVAD-NonConcat and LibriVAD-Concat datasets under the small-scale test setup. As illustrated,  the ViT-MFCC model consistently achieves lowest miss and false acceptance rates among all models, particularly in regions requiring high accuracy, while the ViT-GFCC system performs almost on par. 

\begin{table}
\caption{\it Comparison of VAD performance on the out-of-distribution VOiCES devset using the ViT-MFCC model. Models are trained on the LibriVAD-NonConcat and LibriVAD-Concat datasets, with VAD labels generated by the Montreal Forced Aligner.}
\vspace*{-0.2cm}
\begin{subtable}{0.99\textwidth}
\caption{Models trained on LibriVAD-NonConcat and evaluated on the VOiCES dataset.}
\vspace*{-0.4cm}
\begin{center}
\footnotesize{
\begin{tabular}{|lcccccc|c|c|}\cline{1-9} 
Train data                & Microphone   &Recording  & \multicolumn{4}{|c}{Distractors}        & Avg.     & Overall\\
                    Size &Id (position) &Room       & Babble  & Music  & Television  & None     &  AUC     & Avg. AUC  \\ \hline  \hline
Small                    &clo           & rm1       &.9374 & .9405 & .9390 & .9454 & .9406  &           \\
                         &              & rm2       &.9371 & .9472 & .9419 & .9522 & .9446  &    \\
                         &              & rm3       &.9105 & .8547 & .9074 & .9311 & .9009  &     \\
                         &              & rm4       &.7697 & .9509 & .9458 & .9505 & .9042  & .9225      \\ \cline{2-9}
                         &far           & rm1       &.9021 & .9135 & .9056 & .9291 & .9126  &     \\
                         &              & rm2       &.8990 & .9142 & .9058 & .9354 & .9136  &   \\
                         &              & rm3       &.8007 & .7381 & .7833 & .8717 & .7984  & \\
                         &              & rm4       &.6229 & .8323 & .8318 & .9127 & .7999  &  .8561\\ \hline 
Medium                   & clo          & rm1       &.9709 & .9669 & .9643 & .9659 & .9670  & \\
                         &              & rm2       &.9733 & .9726 & .9708 & .9710 & .9719  & \\
                         &              & rm3       &.9455 & .8864 & .9397 & .9545 & .9315  & \\
                         &              & rm4       &.8041 & .9731 & .9712 & .9717 & .9300 &  .9501\\ \cline{2-9}
                         & far          & rm1       &.9495 & .9485 & .9425 & .9502 & .9477 & \\
                         &              &rm2        &.9413 & .9408 & .9349 & .9467 & .9409 & \\
                         &              & rm3       &.8563 & .7812 & .8196 & .8921 & .8373  & \\
                         &              & rm4       &.6717 & .8785 & .8599 & .9295 & .8349  & .8902\\  \hline
Large                    & clo          &rm1        &.9739 & .9725 & .9690 & .9712 & .9717  &  \\
                         &              &rm2        &.9767 & .9796 & .9777 & .9794 & .9784  & \\
                         &              &rm3        &.9528 & .8905 & .9508 & .9646 & .9397  & \\
                         &              &rm4        &.7983 & .9808 & .9788 & .9799 & .9345  & {\bf .9560} \\ \cline{2-9}
                         & far          &rm1        &.9536 & .9568 & .9462 & .9585 & .9538  &  \\
                         &              &rm2        &.9457 & .9535 & .9431 & .9591 & .9503  &  \\
                         &              &rm3        &.8611 & .7914 & .8319 & .9090 & .8483  & \\
                        &               &rm4         &.6435 & .8897 & .8695 & .9431 & .8364  & {\bf .8972}  \\ \hline
 \end{tabular}
}
\end{center}
\end{subtable}\\[1.5ex]
\begin{subtable}{0.99\textwidth}
\caption{Models trained on LibriVAD-Concat and evaluated on the VOiCES dataset.}
\vspace*{-0.4cm}
\begin{center}
\footnotesize{
\begin{tabular}{|lcccccc|c|c|}\cline{1-9} 
Train data                & Microphone   &Recording  & \multicolumn{4}{|c}{Distractors}        & Avg.     & Overall\\
Size                      &Id (position) &Room       & Babble  & Music  & Television  & None   &  AUC     & Avg. AUC  \\ \hline   \hline   
Small                     & clo          &rm1        & .9536 & .9552 & .9562 & .9602 & .9563 &\\
                          &              &rm2        &.9479 & .9556 & .9568 & .9617 & .9555   &  \\
                          &              &rm3        &.9144 & .8611 & .9156 & .9418 & .9082  & \\
                          &              &rm4        &.8052 & .9620 & .9614 & .9654 & .9235  & .9358 \\ \cline{2-9}
                          & far          &rm1        &.9094 & .9186 & .9180 & .9380 & .9210 & \\
                          &              &rm2        &.8934 & .9000 & .9053 & .9304 & .9073 & \\
                          &              &rm3        &.7970 & .7308 & .7778 & .8701 & .7939 & \\
                          &              &rm4        & .6653 & .8234 & .8484 & .9147 & .8130 & .8588 \\ \hline
Medium                    & clo          & rm1       &.9712 & .9694 & .9687 & .9713 & .9702  & \\
                          &              &rm2        &.9696 & .9730 & .9724 & .9751 & .9725  &\\
                          &              &rm3        &.9466 & .8963 & .9436 & .9614 & .9370  & \\
                          &              &rm4        &.8238 & .9761 & .9753 & .9760 & .9378  & {\bf .9543} \\ \cline{2-9}
                          & far          &rm1        &.9462 & .9514 & .9456 & .9599 & .9508 & \\
                          &              &rm2        &.9366 & .9428 & .9365 & .9548 & .9427 & \\ 
                          &              &rm3        &.8619 & .7931 & .8239 & .9073 & .8465 & \\
                          &              &rm4        &.6818 & .8822 & .8825 & .9454 & .8480 & {\bf .8970} \\ \hline 
Large                     & clo          &rm1        &.9619 & .9582 & .9611 & .9644 & .9614  & \\
                          &              &rm2        &.9567 & .9612 & .9621 & .9659 & .9615  & \\
                          &              &rm3        &.9342 & .8806 & .9326 & .9531 & .9251  &  \\
                          &              &rm4        &.8089 & .9636 & .9652 & .9648 & .9256  & .9434 \\ \cline{2-9}
                          & far          &rm1        &.9290 & .9322 & .9330 & .9514 & .9364  & \\
                          &              &rm2        &.9187 & .9226 & .9247 & .9455 & .9279  & \\
                          &              &rm3        &.8350 & .7709 & .8087 & .8948 & .8274  & \\
                          &              &rm4        &.6717 & .8572 & .8680 & .9290 & .8315  & .8808 \\ \hline
 \end{tabular}
}
\end{center}
\end{subtable}
\label{table:MFA-voices}
\end{table}

\begin{table}[!h]
\caption{\it Silero-VAD performance on the out-of-distribution VOiCES devset.}
\vspace*{-0.2cm}
\begin{center}
\footnotesize{
\begin{tabular}{|lcccccc|c|c|}\cline{1-9} 
Train data                & Microphone   &Recording  & \multicolumn{4}{|c}{Distractors}        & Avg.     & Overall\\
                    Size &Id (position) &Room       & Babble  & Music  & Television  & None     &  AUC     & Avg. AUC  \\ \hline  \hline
Baseline & clo & rm1 & .9733 & .9613 & .9675 & .9683 & .9676 &  \\
(Silero) &              & rm2 & .9565 & .9709 & .9722 & .9771 & .9692 &  \\
         &              & rm3 & .8454 & .8204 & .8753 & .9246 & .8664 &  \\
         &              & rm4 & .7599 & .9667 & .9733 & .9772 & .9193 & .9306 \\ \cline{2-9}

& far & rm1 & .8449 & .8857 & .8965 & .9428 & .8925 &  \\
&     & rm2 & .8106 & .8415 & .8550 & .8927 & .8500 &  \\
&     & rm3 & .7598 & .7446 & .7730 & .8694 & .7867 &  \\
&     & rm4 & .5944 & .7985 & .7960 & .8832 & .7680 & .8243 \\ \hline 

\end{tabular}
}
\end{center}
\label{table:silero_voices}
\end{table}

\begin{table}[!ht]
\caption{\it Comparison of VAD system performance based on EER and MinDCF on the VOiCES dataset.}
\vspace*{-0.2cm}
\begin{subtable}{0.99\textwidth}
\caption{Silero-VAD evaluated on the VOiCES dataset.}
\vspace*{-0.4cm}
\begin{center}
\footnotesize{
\begin{tabular}{|l|l|l|c|}\cline{1-4} 
Model          & Microphone   & EER (\%)  & MinDCF \\ \hline 
Silero-VAD     & clo          & 31.27     & .0664    \\
               & far          & 33.86     & .0999    \\ \hline
\end{tabular}
}
\end{center}
\end{subtable}\\[1.5ex]
\begin{subtable}{0.99\textwidth}
\caption{Models trained on LibriVAD-NonConcat and evaluated on the VOiCES dataset.}
\vspace*{-0.4cm}
\begin{center}
\footnotesize{
\begin{tabular}{|l|l|l|l|c|}\cline{1-5} 
Training Size& Model          & Microphone   & EER (\%)  & MinDCF \\ \hline 
Small     & raw-CLDNN      & clo          & 18.80     & .0835   \\
          &                & far          & 29.27     & .0995   \\
          &                &              &           &     \\
          & MFCC-bDNN      & clo          & 20.61     &.0890   \\
          &                & far          & 28.15     & .0995    \\
          &                &              &           &     \\
          & MFCC-ViT       & clo          & 14.85     & .0883       \\
          &                & far          & 21.86     & .0980      \\ \hline
Medium    & MFCC-ViT       & clo          & 11.45     & .0716       \\
          &                & far          & 18.56     & .0935       \\ \hline          
Large     & MFCC-ViT       & clo          & {\bf 10.42}     & .0674       \\
          &                &far          & {\bf 17.64}     & .0933       \\ \hline
 \end{tabular}
 }
\end{center}
\end{subtable}\\[1.5ex]
\begin{subtable}{0.99\textwidth}
\caption{Models trained on LibriVAD-Concat and evaluated on the VOiCES dataset.}
\vspace*{-0.4cm}
\begin{center}
\footnotesize{
\begin{tabular}{|l|l|l|l|c|}\cline{1-5} 
Training Size& Model      & Microphone   & EER (\%)  & MinDCF \\ \hline 
Small     & raw-CLDNN    & clo          & 17.81    & .0821        \\
          &              & far          & 27.90    & .1000        \\
          &              &              &           &     \\
          & MFCC-bDNN    & clo          & 21.36      & .0894   \\
          &              & far          & 28.65      & .0996   \\ 
          &                &              &           &     \\
          & MFCC-ViT     &clo           & 13.51      &  .0749     \\
          &              & far          & 21.88      & .0939        \\ \hline
Medium    & MFCC-ViT     & clo          & {\bf 10.71}     &  .0647     \\
          &              & far          & {\bf 17.63}      &  .0902 \\ \hline          
Large     & MFCC-ViT     & clo          & 12.55     &  .0688     \\
          &              & far          &  19.60    & .0913       \\ \hline
 \end{tabular}
}
\end{center}
\end{subtable}
\label{table:MFA-voices_EER}
\end{table}

\subsection{Evaluation of VAD Performance on the Out-of-Distribution VOiCES Dataset}
Table \ref{table:MFA-voices} shows the VAD performance of the  ViT-MFCC model on the out-of-distribution VOiCES dataset, with training conducted on the LibriVAD-NonConcat and LibriVAD-Concat datasets. For comparison, the performance of the Silero model is reported in Table \ref{table:silero_voices} as a baseline. 


As shown in Table \ref{table:MFA-voices}, larger training datasets generally lead to higher average AUC values for LibriVAD-NonConcat. For LibriVAD-Concat, the model trained on the large-scale dataset achieves a slightly lower AUC than its medium-scale counterpart, although it still outperforms the model trained on the small-scale dataset with a large margin. These results suggest that increasing the training data size from small to medium is beneficial, likely because the additional data capture a broader range of acoustic conditions and thereby improve generalization. However, the benefits of further scaling are less clear. This trend is consistent with the observations from the in-domain experiments reported in Table \ref{table:small-medium-large-data-size-LibriSpeech}, where scaling from medium to large datasets also failed to yield further performance gains.


Moreover, LibriVAD-Concat and LibriVAD-NonConcat yield broadly comparable AUC scores, indicating that the choice between the two training data sets has only a limited impact on out-of-domain. An exception is the small-scale setting, where the model trained on LibriVAD-Concat slightly outperforms its LibriVAD-NonConcat counterpart. It should also be noted that the VOiCES dataset includes both close-talk and far-field recordings with substantial reverberation, together with acoustic conditions that are absent from LibriVAD. Such domain mismatches present a considerably more challenging evaluation scenario.

Table \ref{table:silero_voices} presents the VAD performance on the VOiCES dataset with the Silero  model. All ViT‑MFCC models outperform the Silero model, with one exception: the ViT‑MFCC model trained on the small‑scale LibriVAD‑NonConcat dataset and evaluated using the close‑talking microphone setup, which performs slightly worse than Silero. In contrast, all ViT‑MFCC models trained on the medium- and large‑scale LibriVAD‑NonConcat and LibriVAD‑Concat datasets outperform the Silero model by a substantial margin, highlighting the effectiveness of both the proposed training datasets and the ViT-MFCC architecture.

Table \ref{table:MFA-voices_EER} presents a comparison of VAD performance for Silero (a) and models trained on LibriVAD datasets of different scales (b) and (c), evaluated on the VOiCES dataset using EER and MinDCF. Scores from audio files across different rooms and distractor conditions are pooled for each microphone when computing EER and MinDCF. The results show that the MFCC-ViT model consistently achieves the lowest EER and MinDCF values in the medium- and large-scale configurations, outperforming other variants and the Silero-VAD baseline, consistent with the trends observed using AUC in Tables \ref{table:MFA-voices} \& \ref{table:silero_voices}.

\begin{table}[!h]
\caption{Effect of the SSR on the VAD performance of MFCC-ViT, trained under the LibriVAD-Concat small-scale setup using different SSR values and evaluated on both the LibriVAD-Concat small-scale test set and the VOiCES dataset. SSR values of 0.2 and 0.5 correspond to LibriVAD-NonConcat and LibriVAD-Concat, respectively; the associated results are reproduced from previous tables for comparison.}
\vspace*{-0.2cm}
\begin{subtable}{0.99\textwidth}
\caption{Models trained on LibriVAD-Concat and evaluated on LibriVAD-Concat for different SSR match and mismatch conditions.}
\begin{center}
\footnotesize{
\begin{tabular}{|c|ccc|ccc|}\cline{1-7} 
 SSR value     & \multicolumn{3}{c|}{Small-\{SSR\}-Concat}  &  \multicolumn{3}{c|}{LibriVAD-Concat-Small} \\ 
               &  AUC   & EER (\%) & MinDCF                 &  AUC   & EER (\%) & MinDCF \\  \hline
    0.2        & .9574& 10.40     & .0528                  & .9613   & 8.99 &.0453 \\              
    0.4        & .9702 & 7.83     & .0362                  & .9687  & 7.73   & .0383 \\  
    0.5        & .9723 & 7.19     & .0353                  &  .9723 & 7.19     & .0353  \\
    0.6        & .9749 & 6.90     & .0362                  & .9728  & 6.99   & .0366 \\ 
    0.8        & .9802 & 5.66     & .0282                  &  .9757 & 6.25   & .0316 \\
    1.0        & .9814 & 5.56     & .0273                  &  .9772 & 6.41   & .0303\\ \hline
\end{tabular}
}
\end{center}

\end{subtable}\\[1.5ex]
\begin{subtable}{0.99\textwidth}
\caption{Models trained on LibriVAD-Concat and evaluated on VOiCES.}
\begin{center}
\footnotesize{
\begin{tabular}{|c|ccclcc|}\cline{1-7} 
 SSR value      & \multicolumn{2}{c}{ }     & \multicolumn{2}{c}{VOiCES  }  & \multicolumn{2}{c|}{} \\ 
               &   AUC (Avg.)               &          &EER (\%)  &        &  MinDCF   &  \\  
               & clo      & far             &  clo     & far      &clo      & far        \\ \hline 
    0.2        & .9225    & .8561           & 14.85   & 21.86     & .0883   & .0980        \\              
    0.4        & .9406    & .8789           & 12.49   &19.66      & .0741   & .0918  \\  
    0.5        & .9358    & .8588           & 13.47   &21.50      & .0756   & .0939  \\
    0.6        & .9505    & .8867           & 11.26   &19.02      & .0670   & .0893     \\ 
    0.8        & .9315    & .8583           & 14.37   &22.19      & .0759   & .0960     \\
    1.0        & .9186    & .8412           & 15.97   &23.83      & .0787   & .0970     \\ \hline
\end{tabular}
}
\end{center}
\end{subtable}
\label{table:MFA-voices_EER_SSR}
\end{table}


To investigate the impact of the proportion of silence in the data, we generated a series of training datasets with SSR values ranging from 0.4 to 1.0 and trained separate models on each dataset. The data generation process followed the LibriSpeech-Concat procedure described in Subsection \ref{subsec:ssr}, differing only in the duration of the injected silence, which was adjusted to obtain the desired SSR level. Note that the datasets for values 0.2 and 0.5 correspond to LibriVAD-NonConcat and LibriVAD-Concat, respectively.  
Table \ref{table:MFA-voices_EER_SSR} compares VAD performance on both the out-of-domain VOiCES dataset and the in-domain datasets. In Table \ref{table:MFA-voices_EER_SSR}(a), the placeholder \{SSR\} in Small-\{SSR\}-Concat denotes the SSR level of the dataset. For each SSR value, the corresponding model was evaluated on a test set with the same SSR level. LibriVAD-Concat-Small refers to the small-scale test set of LibriVAD-Concat, corresponding to an SSR value of 0.5. The results show that performance on both Small-\{SSR\}-Concat and LibriVAD-Concat-Small consistently improves as the SSR increases.

As shown in Table \ref{table:MFA-voices_EER_SSR}(b), an SSR value of 0.6 yields the lowest EER and MinDCF as well as the highest AUC on the VOiCES dataset. This result suggests that maintaining an appropriate balance between speech and nonspeech segments in the training data can enhance the model's ability to generalize to unseen acoustic conditions. This finding contrasts with the observations in Table \ref{table:MFA-voices_EER_SSR}(a), where the evaluation was conducted on in-domain data. Taken together, these findings indicate that the choice of SSR affects in-domain performance and out-of-domain generalization differently.

\subsection{Statistical Significance Analysis}
To assess whether the performance differences between models are statistically significant, we conduct hypothesis testing at the condition level (i.e., the 32 matched test conditions in the VOiCES evaluation set). For each condition, we utilize the average AUC of each model and perform pairwise statistical comparisons using the paired Wilcoxon signed-rank test \cite{c4091bd3-d888-3152-8886-c284bf66a93a}, which is appropriate for paired measurements and does not assume normality of the performance differences. The full matrix of p-values is reported in Table~\ref{tab:statistical_significance} to summarize all pairwise comparisons. To control for multiple comparisons, p-values are adjusted using the Holm–Bonferroni correction \cite{29def780-e117-38f0-8afb-edf384af3fad}.

As the table indicates, each approach is progressively better than the previous at a significance level of at least $a = 0.01$, with the Small-NonConcat approach (the ViT-MFCC model trained on the small-scale LibriVAD-NonConcat dataset) compared against Silero being the only exception.

We employ condition-level analysis, which provides an assessment of generalization across acoustic conditions, to avoid conflating sample size with practical relevance. Conducting hypothesis tests directly on frame-level outputs yields an artificially massive sample size (N). In such regimes, statistical tests become overpowered, rendering even negligible performance differences statistically significant ($p \rightarrow$ 0) solely due to the volume of data \cite{lantz2013large}.

\begin{table}[t]
\centering
\caption{Pairwise Wilcoxon signed-rank test $p$-values comparing the out-of-domain generalization of five models: Silero and the four ViT-MFCC models trained on all combinations of small-scale vs. large-scale training and LibriVAD-NonConcat vs. LibriVAD-Concat datasets across the 32 acoustic conditions in the VOiCES dataset. Models are ordered by ascending overall performance. To account for multiple comparisons, $p$-values are adjusted using the Holm-Bonferroni procedure. Values where $p < 0.05$ indicate a statistically significant difference in performance.}
\label{tab:statistical_significance}
\resizebox{\textwidth}{!}{%
\begin{tabular}{l c c c c c}
\toprule
\textbf{Model} & \multicolumn{1}{c}{\textbf{Silero}} & \multicolumn{4}{c}{\textbf{LibriVAD}} \\
\midrule
\textbf{Training Dataset} &  & \textbf{Small-NonConcat} & \textbf{Small-Concat} & \textbf{Large-NonConcat} & \textbf{Large-Concat} \\
\midrule
\textbf{Silero}          &  & 0.124 & 0.005 & $< 0.001$ & $< 0.001$ \\
\textbf{Small-NonConcat} & 0.124 &  & $< 0.001$ & $< 0.001$ & $< 0.001$ \\
\textbf{Small-Concat}    & 0.005 & $< 0.001$ &  & $< 0.001$ & $< 0.001$ \\
\textbf{Large-NonConcat} & $< 0.001$ & $< 0.001$ & $< 0.001$ &  & $< 0.001$ \\
\textbf{Large-Concat}    & $< 0.001$ & $< 0.001$ & $< 0.001$ & $< 0.001$ &  \\
\bottomrule
\end{tabular}%
}
\end{table}

\subsection{Limitations}
In this study, the proposed LibriSpeech dataset is derived from read English speech recordings. Thus, the developed dataset and systems do not consider the spontaneous speech or languages other than English. Moreover, variations in accents, dialects, and English speech outside the LibriSpeech domain may introduce additional errors in VAD performance.

Although the proposed LibriVAD dataset generation framework primarily focuses on additive noise conditions, the evaluation experiments on the VOiCES dataset additionally assess model robustness under realistic far-field and reverberant environments. Nevertheless, incorporating explicit reverberation simulation and room impulse response modeling into the dataset generation pipeline remains an important direction for future work.

In addition, this work emphasizes deep learning-based benchmarks. However, LibriVAD is a general-purpose dataset with standard waveform inputs and frame-level labels, making it equally applicable to classical, statistical, and hybrid VAD approaches.

\section{Conclusion} \label{Conclusion}
In this work, we address a critical bottleneck in advancing robust voice activity detection (VAD): the scarcity of large-scale, systematically controlled, and publicly available datasets for training and evaluation. To overcome this, we introduced \textbf{LibriVAD}, a scalable open-source dataset meticulously constructed from the LibriSpeech corpus and enriched with a diverse array of real-world and synthetic noise sources. 
LibriVAD offers fine-grained control over key variables such as SNR and SSR, and is available in three sizes (15 GB, 150 GB, and 1.5 TB), making it a flexible resource for rigorously benchmarking VAD systems under a wide spectrum of challenging acoustic conditions.

Our extensive experiments establish strong baselines for deep learning-based VAD models and demonstrate that the Vision Transformer architecture combined with MFCC features consistently outperforms conventional approaches.  
This configuration not only achieves superior performance under both seen and unseen conditions within LibriVAD but also generalizes effectively to out-of-distribution real-world data from the VOiCES dataset, underscoring its robustness and potential as a high-performing approach for VAD. 

Furthermore, our findings show that scaling the dataset from a small-scale setting to medium- or large-scale settings substantially improves model generalization, particularly in out-of-distribution, real-world scenarios. Evaluation on the VOiCES dataset further demonstrates the robustness of the proposed approach in reverberant and noisy environments.

By publicly releasing the LibriVAD dataset, data generation pipeline, baseline implementations, and trained models, we aim to promote reproducibility and accelerate advancements in VAD research. We envision LibriVAD as a comprehensive benchmark that will drive the development of innovative VAD technologies. Future work will extend LibriVAD to include multiple languages, accents, and speech styles, enabling more inclusive and globally applicable VAD systems.

\vspace{12pt}
\textbf{Declaration of generative AI and AI-assisted technologies in the manuscript preparation process}

During the preparation of this work the authors used ChatGPT in order to improve clarity and English language usage at the sentence level. After using this tool/service, the authors reviewed and edited the content as needed and take full responsibility for the content of the published article.

\vspace{12pt}
\textbf{Acknowledgements}
I. Stylianou and Z.-H. Tan are supported by the Pioneer Centre for AI, Denmark. This collaboration was initiated during Tan’s visit to MIT Computer Science and Artificial Intelligence Laboratory, USA. A. K. Sarkar is supported by the NLTM BHASHINI project funding (11(1)/2022-HCC(TDIL)) from the Ministry of Electronics and Information Technology (MeitY), Government of India.

\appendix

\section{Dataset Structure}
The LibriVAD dataset is distributed as a set of generation scripts and metadata hosted on Hugging Face. To facilitate reproducibility and ease of use, we outline the structure of the repository and the resulting directory hierarchy of the generated audio data below.

\subsection{Hugging Face Repository Structure}
The repository (\url{https://huggingface.co/datasets/LibriVAD/LibriVAD}) contains the necessary Python scripts to download the source LibriSpeech data, augment it with specific noise types, and generate the corresponding VAD labels.

\dirtree{%
.1 LibriVAD/.
.2 create\_LibriVAD.py\ldots{} \textit{Main script for dataset generation}.
.2 setup.py\ldots{} \textit{Script for downloading LibriSpeech, setting up noises, generating silence etc}.
.2 requirements.txt\ldots{} \textit{Python dependencies}.
.2 README.md\ldots{} \textit{Documentation}.
.2 Files/.
.3 Noises.zip\ldots{} \textit{Source noise files}.
.4 Babble\_noise/\ldots{} \textit{WAV noisy files}.
.4 Domestic\_noise/.
.4 \textit{[Other Noise Types]...}.
.3 Forced\_alignments.zip.
.4 librispeech\_alignments/.
.5 train-clean-100/\ldots{} \textit{.TextGrid labels}.
.5 dev-clean/.
.5 test-clean/.
.5 unaligned.txt\ldots{} \textit{Files without forced alignments}.
.3 Datasets/\ldots{} \textit{Destination for clean audio downloads}.
}

\subsection{Generated Dataset Directory Structure}
Upon executing the generation scripts, the dataset is organized hierarchically by dataset split, noise type, and signal-to-noise ratio. This structure allows for easy data loading and supports stratified sampling.

\dirtree{%
.1 LibriVAD/.
.2 Results/.
.3 LibriSpeech/.
.4 train-clean-100/.
.5 Babble\_noise/.
.6 -5/\ldots{} \textit{WAV files at -5dB SNR}.
.6 0/.
.6 5/.
.6 10/.
.6 15/.
.6 20/.
.5 Domestic\_noise/.
.5 \textit{[Other Noise Types]...}.
.4 dev-clean/.
.5 \textit{[Same structure as train-clean-100]}.
.4 test-clean/.
.5 \textit{[Same structure as train-clean-100]}.
.3 LibriSpeechConcat/.
.4 \textit{[Same structure as LibriSpeech]}.
}

\subsection{Label Format}
For every generated audio file (e.g., \texttt{audio\_1.wav}), a corresponding label file (e.g., \texttt{audio\_1.npy}) is provided in the \texttt{Files/Labels/} directory. These labels represent frame-level binary vectors where $1$ denotes speech activity and $0$ denotes non-speech or silence.

\bibliographystyle{elsarticle-num}
\bibliography{references}

@inproceedings{dosovitskiy2020image,
  title={An Image is Worth 16x16 Words: Transformers for Image Recognition at Scale},
  author={Dosovitskiy, Alexey and Beyer, Lucas and Kolesnikov, Alexander and Weissenborn, Dirk and Zhai, Xiaohua and Unterthiner, Thomas and Dehghani, Mostafa and Minderer, Matthias and Heigold, G and Gelly, S and others},
  booktitle={International Conference on Learning Representations},
  year={2020}
}

@article{ding2019personal,
  title={Personal VAD: Speaker-conditioned voice activity detection},
  author={Ding, Shaojin and Wang, Quan and Chang, Shuo-yiin and Wan, Li and Moreno, Ignacio Lopez},
  journal={arXiv preprint arXiv:1908.04284},
  year={2019}
}

@inproceedings{bovbjerg2024self,
  title={Self-supervised pretraining for robust personalized voice activity detection in adverse conditions},
  author={Bovbjerg, Holger Severin and Jensen, Jesper and {\O}stergaard, Jan and Tan, Zheng-Hua},
  booktitle={ICASSP 2024-2024 IEEE International Conference on Acoustics, Speech and Signal Processing (ICASSP)},
  pages={10126--10130},
  year={2024},
  organization={IEEE}
}

@inproceedings{larsen2022adversarial,
  title={Adversarial Multi-Task Deep Learning for Noise-Robust Voice Activity Detection with Low Algorithmic Delay},
  author={Larsen, Claus Meyer and Koch, Peter and Tan, Zheng-Hua},
  booktitle={Interspeech 2022},
  pages={3759--3763},
  year={2022}
}

@article{hoang2022minimum,
  title={The Minimum Overlap-Gap Algorithm for Speech Enhancement},
  author={Hoang, Poul and Tan, Zheng-Hua and De Haan, Jan Mark and Jensen, Jesper},
  journal={IEEE Access},
  volume={10},
  pages={14698--14716},
  year={2022},
  publisher={IEEE}
}

@Article{Sohn1999a,
author={J. Sohn and N. S. Kim and W. Sung},
title={{A} statistical model-based voice activity detection},
journal={IEEE Signal Processing Letters},
volume={6},
number = {1},
pages={1-3},
year={1999}
}

@inproceedings{Hirsch2000,
author ={ H.-G. Hirsch and D. Pearce},
title = {{T}HE AURORA EXPERIMENTAL FRAMEWORK FOR THE PERFORMANCE EVALUATION OF SPEECH RECOGNITION SYSTEMS UNDER NOISY CONDITIONS},
booktitle ={Automatic Speech Recognition: Challenges for the Next Millennium, ISCA ITRW ASR2000},
year= {2000}
}

@inproceedings{hansen19_interspeech,
  author={J. H.L. Hansen and A. Joglekar and M. C. Shekhar and V. Kothapally and C. Yu and L. Kaushik and A. Sangwan},
  title={{The 2019 Inaugural Fearless Steps Challenge: A Giant Leap for Naturalistic Audio}},
  year=2019,
  booktitle={Proc. Interspeech},
  pages={1851-1855}
 }

@INPROCEEDINGS{kraljevski2015comparison,
  title={Comparison of forced-alignment speech recognition and humans for generating reference VAD},
  author={Kraljevski, Ivan and Tan, Zheng-Hua and Bissiri, Maria Paola},
  booktitle ={Proc. of Interspeech},
  year={2015},
  pages= {2937-941}
  }

@Article{Davis80,
	author={S. B. Davis and P. Mermelstein},
	title={{C}omparison of parametric representations for monosyllabic word recognition in continuously spoken sentences},
	journal={IEEE Trans. Acoust. Speech Signal Processing},
	volume={28},
	pages={357-366},
	year={1980}
}

@INPROCEEDINGS{Wave_2020,
  author = {Yu, C. and Hung, K.-H. and Lin, I-F. and Fu, S.-W. and Tsao, Y. and Hung, J.-W.},
  title = {{W}aveform-based Voice Activity Detection Exploiting Fully Convolutional networks with Multi-Branched Encoders},
   booktitle= {arXiv},
  year = {2020}
  }

@ARTICLE{7347379,  
author={Zhang, X.-L. and Wang, D.}, 
journal={IEEE/ACM Transactions on Audio, Speech, and Language Processing}, 
title={Boosting Contextual Information for Deep Neural Network Based Voice Activity Detection},  
year={2016},  volume={24},  number={2},  pages={252-264}
}

@article{cosentino2020librimix,
  title={Librimix: An open-source dataset for generalizable speech separation},
  author={Cosentino, Joris and Pariente, Manuel and Cornell, Samuele and Deleforge, Antoine and Vincent, Emmanuel},
  journal={arXiv preprint arXiv:2005.11262},
  year={2020}
}

@article{wichern2019wham,
  title={Wham!: Extending speech separation to noisy environments},
  author={Wichern, Gordon and Antognini, Joe and Flynn, Michael and Zhu, Licheng Richard and McQuinn, Emmett and Crow, Dwight and Manilow, Ethan and Roux, Jonathan Le},
  journal={arXiv preprint arXiv:1907.01160},
  year={2019}
}

@article{khan2022transformers,
  title={Transformers in vision: A survey},
  author={Khan, Salman and Naseer, Muzammal and Hayat, Munawar and Zamir, Syed Waqas and Khan, Fahad Shahbaz and Shah, Mubarak},
  journal={ACM computing surveys (CSUR)},
  volume={54},
  number={10s},
  pages={1--41},
  year={2022},
  publisher={ACM New York, NY}
}

@article{tan2020rvad,
title = {{rVAD}: An unsupervised segment-based robust voice activity detection method},
journal = {Computer Speech \& Language},
volume = {59},
pages = {1-21},
year = {2020},
author = {Z.-H. Tan and A. K. Sarkar and N. Dehak},
note    = {Available at: \url{https://github.com/zhenghuatan/rVAD}.}
}

@inproceedings{zazo2016feature,
  title={Feature Learning with Raw-Waveform CLDNNs for Voice Activity Detection.},
  author={Zazo, Rub{\'e}n and Sainath, Tara N and Simko, Gabor and Parada, Carolina},
  booktitle={Interspeech},
  pages={3668--3672},
  year={2016}
}

@article{li2024robust,
  title={Robust voice activity detection using an auditory-inspired masked modulation encoder based convolutional attention network},
  author={Li, Nan and Wang, Longbiao and Ge, Meng and Unoki, Masashi and Li, Sheng and Dang, Jianwu},
  journal={Speech Communication},
  volume={157},
  pages={103024},
  year={2024},
  publisher={Elsevier}
}

@dataset{thiemann_2018_1227121,
  author       = {Thiemann, Joachim and
                  Ito, Nobutaka and
                  Vincent, Emmanuel},
  title        = {{DEMAND: a collection of multi-channel recordings 
                   of acoustic noise in diverse environments}},
  month        = apr,
  year         = 2018,
  publisher    = {Zenodo},
  version      = {1.0},
  doi          = {10.5281/zenodo.1227121},
}

@inproceedings{panayotov2015librispeech,
  title={Librispeech: an asr corpus based on public domain audio books},
  author={Panayotov, Vassil and Chen, Guoguo and Povey, Daniel and Khudanpur, Sanjeev},
  booktitle={2015 IEEE international conference on acoustics, speech and signal processing (ICASSP)},
  pages={5206--5210},
  year={2015},
  organization={IEEE}
}

@inproceedings{kahn2020libri,
  title={Libri-light: A benchmark for asr with limited or no supervision},
  author={Kahn, Jacob and Riviere, Morgane and Zheng, Weiyi and Kharitonov, Evgeny and Xu, Qiantong and Mazar{\'e}, Pierre-Emmanuel and Karadayi, Julien and Liptchinsky, Vitaliy and Collobert, Ronan and Fuegen, Christian and others},
  booktitle={ICASSP 2020-2020 IEEE International Conference on Acoustics, Speech and Signal Processing (ICASSP)},
  pages={7669--7673},
  year={2020},
  organization={IEEE}
}

@inproceedings{Berg_2021, 
   title={Keyword Transformer: A Self-Attention Model for Keyword Spotting},
   booktitle={Proc of Interspeech},
   author={Berg, A. and O’Connor, M. and Cruz, M. T.},
   year={2021},
   pages={4249--4253}
}

@inproceedings{Fujita2019,
author={Y. Fujita and  N. Kanda and  S. Horiguchi and  K. Nagamatsu and  S. Watanabe},
title={End-to-end neural speaker diarization with
permutation-free objectives}, 
booktitle ={Proc of Interspeech},
year = {2019}, 
pages = {4300-4304}
}

@inproceedings{Horiguchi2020,
author={S. Horiguchi and Y. Fujita and S. Watanabe and  Y. Xue and  K. Nagamatsu},
title = {End-to-end speaker diarization for an unknown
number of speakers with encoder-decoder based attractors},
booktitle ={Proc of Interspeech},
year={2020},
pages = {269-273}
}

@INPROCEEDINGS{10890295,
  author={Z. Chen and B. Han and S. Wang and Y. Jiang and Y. Qian},
  booktitle={Proc. of IEEE Int. Conf. Acoust. Speech Signal Processing (ICASSP)},
  title={Flow-TSVAD: Target-Speaker Voice Activity Detection via Latent Flow Matching for Speaker Diarization}, 
  year={2025},
  pages={1-5}
  }

@INPROCEEDINGS{sarkar2023,
author = {A. K. Sarkar and others}, 
title = {Study of Various End-to-End Keyword Spotting Systems on the Bengali Language Under Low-Resource Condition},
booktitle = {Springer Nature Switzerland},
year = {2023},
pages = {114-126}
}

@INPROCEEDINGS{10890730,
  author={B. Labrador et al},
  booktitle={Proc. of IEEE Int. Conf. Acoust. Speech Signal Processing (ICASSP)}, 
  title={Personalizing Keyword Spotting with Speaker Information}, 
  year={2025},
  pages={1-5}
  }

@ARTICLE{10858118,
  author={Z. -E. Wu, S. -J. Chan, Y. A. Wubet and K. -Y. Lian},
  journal={IEEE Access}, 
  title={DLiGRU-X: Efficient X-Vector-Based Embeddings for Small-Footprint Keyword Spotting System}, 
  year={2025},
  volume={13},
  pages={23498-23507}
  }

@ARTICLE{Opochinsky2025,
author = {Opochinsky, R. and Moradi, M. and Gannot, S.},
title = {Single-microphone speaker separation and voice activity detection in noisy and reverberant environments},
journal = { J AUDIO SPEECH MUSIC PROC.},
year = {2025},
volume = {18}
}

@INPROCEEDINGS{10888445,
  author={T. Cord-Landwehr and C. Boeddeker and R. Haeb-Umbach},
  booktitle={Proc. of IEEE Int. Conf. Acoust. Speech Signal Processing (ICASSP)}, 
  title={Simultaneous Diarization and Separation of Meetings through the Integration of Statistical Mixture Models}, 
  year={2025},
  pages={1-5}
  }

@article{HOANG2025104969,
title = {Lightweight speaker verification with integrated VAD and speech enhancement},
journal = {Digital Signal Processing},
volume = {159},
pages = {104969},
year = {2025},
issn = {1051-2004},
author = {K. A. Hoang and T. Le and H. T. Nguyen}
}

@INPROCEEDINGS{10023187,
  author={Liu, Alexander H. and Hsu, Wei-Ning and Auli, Michael and Baevski, Alexei},
  booktitle={IEEE Spoken Language Technology Workshop (SLT)}, 
  title={Towards End-to-End Unsupervised Speech Recognition}, 
  year={2023},
  pages={221-228}
  }

@INPROCEEDINGS{9747357,
  author={G. -T. Lin and C. -J. Hsu and D. -R. Liu and H. -Y. Lee and Y. Tsao},
  booktitle={Proc. of IEEE Int. Conf. Acoust. Speech Signal Processing (ICASSP)},  
  title={Analyzing The Robustness of Unsupervised Speech Recognition}, 
  year={2022},
  pages={8202-8206},
  }

@misc{SRE,
           author = {NIST},
           title = {NIST Speaker Recognition Evaluation ({SRE}) series},
           howpublished = {\url{https://www.nist.gov/itl/iad/mig/speaker-recognition}},
           year = {2025},
           note = {Accessed on 2025-06-29}
       }

@misc{LRE,
           author = {NIST},
           title = {NIST Language Recognition Evaluation ({LRE}) series },
           howpublished = {\url{https://www.nist.gov/itl/iad/mig/language-recognition}},
           year = {2025},
           note = {Accessed on 2025-06-29}
       }

@inproceedings{piczak2015dataset,
  title = {{ESC}: Dataset for Environmental Sound Classifications},
  author = {Piczak, Karol J.},
  booktitle = {Proc. of the 23rd Annual ACM Conference on Multimedia},
  year = {2015},
  pages = {1015--1018}
}

@INPROCEEDINGS{heittola2020acousticsceneclassificationdcase,
      title={Acoustic scene classification in DCASE 2020 Challenge: generalization across devices and low complexity solutions}, 
      author={T. Heittola and A. Mesaros and T. Virtanen},
      booktitle = {Proc. of the Detection and Classification of Acoustic Scenes and Events 2020 Workshop (DCASE2020)},
      year={2020}
    }

@INPROCEEDINGS{Timit,
author = {J. S. Garofolo and L. F. Lamel and W. M. Fisher and J. G. Fiscus and  D. S. Pallett and N. L. Dahlgren and V. Zue},
title = {{TIMIT} acoustic-phonetic continuous speech corpus LDC93S1},
year={1993}
}

@INPROCEEDINGS{Walker2012,
author ={K. Walker and S. Strassel},
title={{T}he RATS radio traffic collection system},
booktitle={Proc. of Odyssey Speaker and Language Recognition Workshop},
year={2012},
pages= {291-297}
}

@INPROCEEDINGS{4517928,
  author={Y. Shao and D. Wang},
  booktitle={Proc. of IEEE Int. Conf. Acoust. Speech Signal Processing (ICASSP)}, 
  title={Robust speaker identification using auditory features and computational auditory scene analysis}, 
  year={2008},
  pages={1589-1592},
  }

@ARTICLE{Valero,
author={Valero, X. and F. Alias},
title={Gammatone Cepstral Coefficients: Biologically Inspired Features for Non-Speech Audio Classification},
journal={IEEE Transactions on Multimedia},
volume={14},
number ={6},
year ={2012},
pages = {1684-1689}
}

@INPROCEEDINGS{fearlessstepschallengephase1,
      title={Fearless Steps Challenge Phase-1 Evaluation Plan}, 
      author={A. Joglekar and J. H. L. Hansen},
      year={2022},
      booktitle={arXiv}      
}

@misc{LDC93S6A,
  title        = {CSR-I (WSJ0) Complete},
  author       = {Garofolo, John S. and Graff, David and Baker, Janet M. and Paul, Doug and Pallett, David},
  howpublished = {Linguistic Data Consortium, LDC93S6A},
  year         = {1993},
  note         = {Web Download, Philadelphia},
}

@INPROCEEDINGS{richey2018voices,
    title={Voices Obscured in Complex Environmental Settings (VOICES) corpus},
    author={C. Richey and M. A. Barrios and Z. Armstrong and C. Bartels and H. Franco and M. Graciarena and A. Lawson and M. K. Nandwana and A. Stauffer and J. V. Hout and P. Gamble and J. Hetherly and C. Stephenson and K. Ni},
    year={2018},
    booktitle={Proc. Interspeech 2018}
    }

@article{yu2017spoofing,
  title={Spoofing detection in automatic speaker verification systems using DNN classifiers and dynamic acoustic features},
  author={Yu, Hong and Tan, Zheng-Hua and Ma, Zhanyu and Martin, Rainer and Guo, Jun},
  journal={IEEE transactions on neural networks and learning systems},
  volume={29},
  number={10},
  pages={4633--4644},
  year={2017},
  publisher={IEEE}
}

@inproceedings{dean10_interspeech,
  title     = {The QUT-NOISE-TIMIT corpus for the evaluation of voice activity detection algorithms},
  author    = {D. Dean and S. Sridharan and R. Vogt and M. Mason},
  year      = {2010},
  booktitle = {Proc of Interspeech},
  pages     = {3110--3113}
  }

@misc{OpenSAD15,
           author = {OpenSAD15},
           title = {Evaluation Plan for the NIST Open Evaluation
of Speech Activity Detection ({OpenSAD15})},
           howpublished = {\url{https://www.nist.gov/system/files/documents/itl/iad/mig/Open_SAD_Eval_Plan_v10.pdf}},
           year = {2015},
           note = {Accessed on 2025-06-29}
       }

@misc{Jemine2017LibrispeechAlignments,
  author    = {Jemine, Corentin},
  title     = {LibriSpeech Alignments},
  year      = {2017},
  publisher = {GitHub},
  journal   = {GitHub repository},
  howpublished = {\url{https://github.com/CorentinJ/librispeech-alignments}},
  note      = {Accessed: 2025-11-06}
}

@inproceedings{mcauliffe17_interspeech,
  title     = {Montreal Forced Aligner: Trainable Text-Speech Alignment Using Kaldi},
  author    = {M. McAuliffe and M. Socolof and S. Mihuc and M. Wagner and M. Sonderegger},
  year      = {2017},
  booktitle = {Proc of Interspeech},
  pages     = {498--502}
  }

@software{smith2024mycode,
  author       = {W. J. Jose},
  title        = {Keyword-transformer},
  year         = {2021},
  url          = {https://github.com/wdjose/keyword-transformer}
}

@article{goldman1968psycholinguistics,
  title={Psycholinguistics: Experiments in spontaneous speech},
  author={Goldman-Eisler, Frieda},
  year={1968},
  publisher={Academic Press}
}

@article{henderson1966sequential,
  title={Sequential temporal patterns in spontaneous speech},
  author={Henderson, Alan and Goldman-Eisler, Frieda and Skarbek, Andrew},
  journal={Language and Speech},
  volume={9},
  number={4},
  pages={207--216},
  year={1966},
  publisher={Sage Publications Sage CA: Thousand Oaks, CA}
}

@misc{SileroVAD,
  author = {Silero Team},
  title = {Silero VAD: pre-trained enterprise-grade Voice Activity Detector (VAD), Number Detector and Language Classifier},
  year = {2024},
  publisher = {GitHub},
  journal = {GitHub repository},
  howpublished = {\url{https://github.com/snakers4/silero-vad}}
 
}

@Article{tsne,
author={L. J. P. van der Maaten and G. E. Hinton},
title={{V}isualizing {H}igh-{D}imensional {D}ata {U}sing {t-SNE}},
journal={Journal of Machine Learning Research},
volume={9},
pages={2579-2605},
year={2008}
}

@inproceedings{chaudhuri18_interspeech,
  title     = {{AVA-Speech: A Densely Labeled Dataset of Speech Activity in Movies}},
  author    = {Sourish Chaudhuri and Joseph Roth and Daniel P. W. Ellis and Andrew Gallagher and Liat Kaver and Radhika Marvin and Caroline Pantofaru and Nathan Reale and Loretta {Guarino Reid} and Kevin Wilson and Zhonghua Xi},
  year      = {2018},
  booktitle = {Proc of Interspeech},
  pages     = {1239-1243}
  
}

@INPROCEEDINGS{NTimit,
author = {W. M. Fisher and G. R. Doddington and K. M. Goudie-Marshall and C. Jankowski and A. Kalyanswamy and S. Basson and J. Spitz},
title = {{NTIMIT}  {LDC93S2} },
booktitle = {Web Download. Philadelphia: Linguistic Data Consortium},
year={1993}
}

@article{c4091bd3-d888-3152-8886-c284bf66a93a,
 ISSN = {00994987},
 URL = {http://www.jstor.org/stable/3001968},
 author = {Frank Wilcoxon},
 journal = {Biometrics Bulletin},
 number = {6},
 pages = {80--83},
 publisher = {[International Biometric Society, Wiley]},
 title = {Individual Comparisons by Ranking Methods},
 urldate = {2026-02-26},
 volume = {1},
 year = {1945}
}

@article{29def780-e117-38f0-8afb-edf384af3fad,
 ISSN = {03036898, 14679469},
 URL = {http://www.jstor.org/stable/4615733},
 author = {Sture Holm},
 journal = {Scandinavian Journal of Statistics},
 number = {2},
 pages = {65--70},
 publisher = {[Board of the Foundation of the Scandinavian Journal of Statistics, Wiley]},
 title = {A Simple Sequentially Rejective Multiple Test Procedure},
 urldate = {2026-02-26},
 volume = {6},
 year = {1979}
}

@article{lantz2013large,
  title={The large sample size fallacy},
  author={Lantz, Bj{\"o}rn},
  journal={Scandinavian journal of caring sciences},
  volume={27},
  number={2},
  pages={487--492},
  year={2013},
  publisher={Wiley Online Library}
}
\end{document}